\documentclass[preprint2]{aastex}%
\usepackage{url}

\usepackage{amsmath}
\usepackage{color}

\RequirePackage{color}

\begin{document}

\title{Dynamical Monte Carlo simulations of 3-D galactic systems in axisymmetric and triaxial potentials}

\shorttitle{Dynamical Monte Carlo simulations of 3-D galactic systems in axisymmetric and triaxial potentials}
\shortauthors{A. Taani \& J. C. Vallejo}

\author{Ali Taani}
\affil{ Applied Science Department, Aqaba University College, Al-Balqa Applied
University, P.O. Box 1199, Aqaba, Jordan}
\email{ali82taani@gmail.com}

\author{Juan C. Vallejo}
\affil{ European Space Astronomy Centre, PO Box 78, E-28691 Villanueva de la Canada, Madrid, Spain}


\begin{abstract}
We describe the dynamical behavior of isolated old ($\geq 1$
Gyr) objects-like Neutron Stars (NSs). These isolated NSs
are evolved under smooth, time-independent, 3-D gravitational potentials, axisymmetric and with a triaxial dark halo. We analysed
the geometry of the dynamics and applied the Poincar\'{e} section method for comparing the influence of different birth velocity distributions.
The inspection of the maximal asymptotic Lyapunov ($\lambda$) exponent shows that the dynamical behaviors of the selected
orbits are nearly the same as the regular orbits with two degrees of freedom, both in axisymmetric and
triaxial when ($\phi$, $q_{z}$)= (0,0).  Conversely, a few chaotic trajectories are found
with a rotated triaxial halo when ($\phi$, $q_{z}$)= (90, 1.5). The tube orbits preserve the direction of their circulation around either the
long or short axis as appeared in the triaxial potential, even when every initial condition leads to
different orientations. The Poincar\'{e} section method shows that there are 2-D invariant tori
and invariant curves (islands) around stable periodic orbits that bound to the surface of 3-D tori.
The regularity of the several prototypical orbits offer the means to identify the phase space regions
with localized motions and to determine their environment in different models, because they can occupy significant parts of
phase-space depending on the potential. This is of particular importance in Galactic Dynamics.
\end{abstract}


\section{Introduction}
Neutron stars (NSs) manifest the most extreme values of many stellar and physical
parameters including spin period,
orbital parameters, magnetic field and kick velocity.
These are typically old systems (1-10 Gyr) and relatively short timescales
($\sim$ 10$^{6}$ yr) (e.g.,
Yakovlev \& Pethick 2004; Lorimer 2008) and show a concentration towards the
Galactic center bulge (which is at $\sim$ 8 kpc) and in the globular clusters
(Caranicolas \& Zotos 2009; Taani et al. 2012a; Kalamkar 2013; Taani 2016).

It is generally believed that Millisecond Pulsar (MSPs) are very old NSs spun up owing
to mass accretion during
the phase of mass exchange in binaries (Alpar et al. 1982). These systems are detectable
as active radio pulsars only when they recycled (spun up).
While the isolated old NSs  have not been identified (Haberl 2007; Kaplan 2008), and we
define them here by their steady flux,
predominantly thermal X-ray emission, lack of optical or radio counterparts, and the
absence of a surrounding pulsar wind nebula
(e. g. Ostriker et al. 1970;  Shvartsman 1971; Neuh$\ddot{a}$user \& Tr\"{u}mper 1999;
Treves et al. 2000; Pavlov et al. 2004;
De Luca 2008; Halpern \& Gotthelf 2010). As a consequence
little is known about their physical and statistical properties. One would hope that the
isolated old NSs may be detected as soft X-ray
sources (0.5-2 keV) in the $\emph{eRosita}$ all-sky survey (Merloni et al.
2012; Doroshenko et al. 2014). However, the estimation of the pulsar velocities
depends on the direct distance measurements (Hobbs et al. 2005) which can be obtained
by the dispersion measure and a Galactic density
model (Paczy\'{n}ski 1990; Hansen \& Phinney 1997; Cordes \& Chernoff 1998; Lorimer 2008;
 Sartore et al. 2010). These studies give a
mean birth velocity $100$ - $500$ km s$^{-1}$, with possibly a significant
population having $\emph{v}$ $\geq$ 1000 km $s^{-1}$. Arzoumanian et al. (2002)
favor a bimodal pulsar velocity distribution, with peaks around
100 and 500 km $s^{-1}$. However the mechanisms of high velocity are still
open questions (Wang \& Han 2010).

Isolated old NSs have attracted much attention because of the
hope that their properties could be used to constrain the poorly
understood behavior. Studying their orbital dynamics
in known gravitational potential is very significant to our understanding
the Galactic gravitational field, as well as the evolution of the
Galactic disk structure it self. Paczy\'{n}ski (1990) (hereafter P90) simulated
the motion of NSs in a galactic potential and calculated the NS space
density distribution. In the same work P90 also suggested
a simplified expression for the gravitational potential
which is still often applied in the simulation of NS distribution.


This paper is the third in a series of papers. Wei et al. (2010a, hereafter Paper I)
used this potential, and they also adopted a Galactic
distribution with one-component initial random velocity models. Wei et al. (2010b)
investigated the above gravitational potential
of the Galactic disk and several prototypical orbits of stars, and found that all of
the orbits are symmetric with respect to the galactic plane.
Taani et al. (2012b, hereafter Paper II) constructed a phenomenological model
with the same gravitational potential, but under a two-component
Maxwellian initial random velocity distribution following P90 and
Faucher-Gigu µe~re \& Kaspi (2006). The conclusion was that
there are some non-symmetric orbits. When the motion
ranges in the vertical direction and hence becomes larger than the
one in the radial direction,
the orbits become more regular. It was also found that
the irregular character of the motion of NSs increases when the vertical direction
becomes larger than radial direction. We mean here when the motion is out of the disk plane.
The large Galactic radial
expansion (understood as $\emph{R} \sim 15 \emph{Kpc}$) could give hints to the distribution
 of NS progenitors. The majority of them (80\%) falls within $\emph{R}\leq$25 \emph{kpc} from the Galactic
rotation axis.

In the present paper we extend and complete the study of the old NS sample,
by describing the dynamical orbital evolution of the isolated old NSs.
These isolated NSs are evolved under a smooth, time-independent, 3-D axisymmetric
gravitational potential that represents a non-rotated galactic disk. In addition, we extend the purpose of the present study to explore the orbital
dynamics of realistic triaxial potential in a systematic way.
We focus on plotting the 3-D trajectories and their 2-D projections under a variety of
initial conditions. These initial conditions are obtained by performing Monte Carlo
simulations to develop perturbation approximations of the 3-D orbits.

Our main objective is to investigate the regular or chaotic nature of the
computed trajectories. It is noteworthy to mention that Evans (1994) found some semi-stochastic orbits in triaxial potentials, because the chaos was tentatively associated with linear instability of the short-and intermediate-axis orbits, while Goodman \& Schwarzschild (1981) discussed the importance of stochastic orbits in triaxial models.

The structure of the paper is as follows. First, we introduce the model and the Monte
Carlo technique we have used. Then, we will analyze the dynamical properties
of a set of selected initial conditions, including the Poincar\'{e} sections, Lyapunov Asymptotic Exponents and dark triaxial haloes case.
Finally, we will end with some conclusions.

\begin{figure}
\includegraphics[angle=0,width=8.0cm]{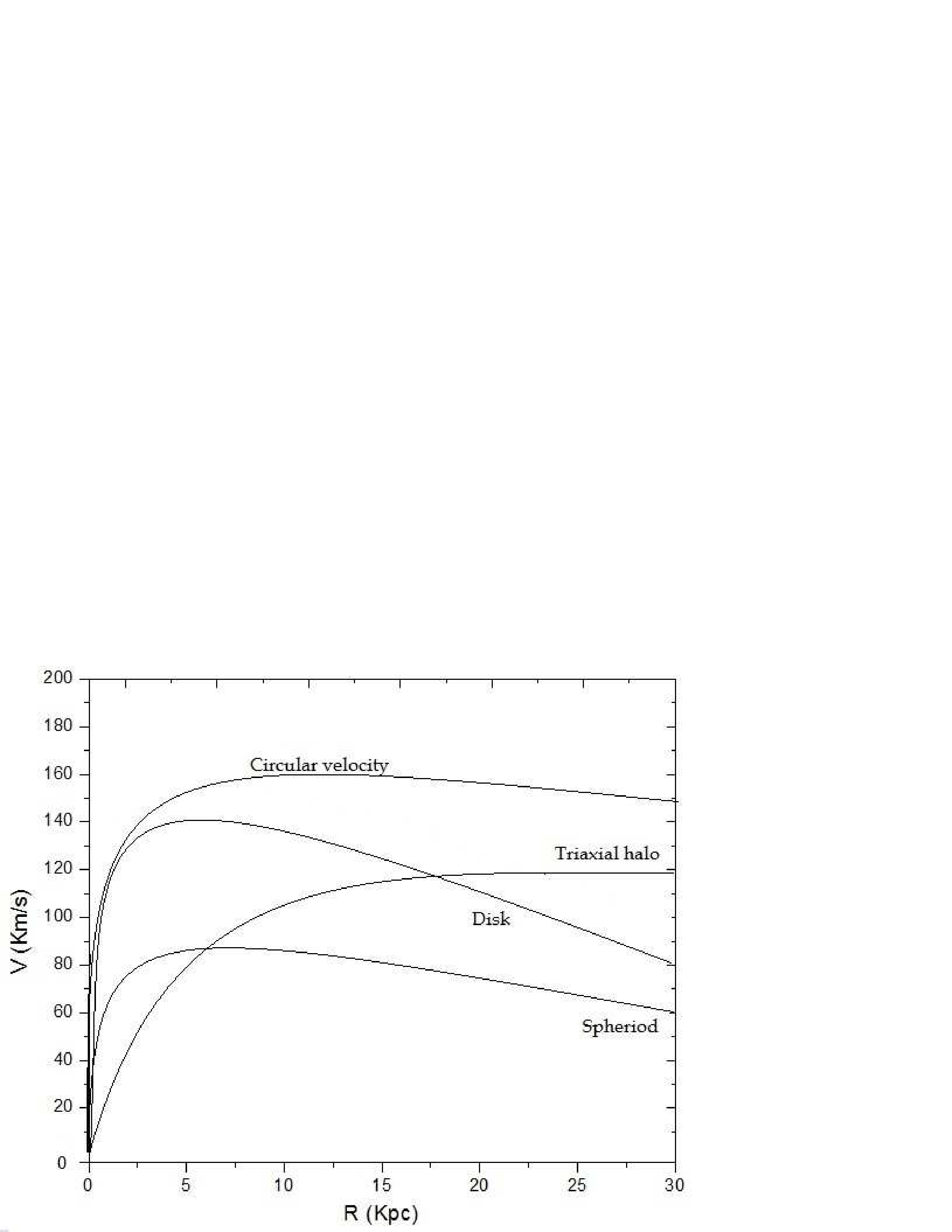}
\caption{ Rotation curve of the axisymmetric potential includes a Miyamoto-Nagai disk and spheriod, triaxial halo and averaged circular velocity in the galactic plane.}
 \label{rotation-curve}
\end{figure}

\section{Simulation and Numerical setup}

\subsection{Galactic Gravitational Potential}

The equations presented in this paper describe how could the motion and position of NSs
be affected by the Galactic gravitational potential.
We will follow  papers I \& II in their procedures to introduce the P90 Galactic
gravitational potential. This model is time-independent and very simple.
The integration of the orbit, using this model, is
very rapid and can achieve high numerical precision.

It is noteworthy to mention here that the P90
model is taken to be a homogeneous function of the
density, and ignore the interstellar friction. This
is a reliable approximation for our axisymmetric
model, because the steady state distribution of old
NSs depends only weakly on the non-homogeneous
part of the galactic potential (Frei et al. 1992).
Using P90 may not be a good approximation when
studying non-axisymmetric models, because rotating
non-axisymmetric components (like bars or
spirals) can introduce resonances (Patsis et al. 2002).
This model is time-independent and very simple. The integration of the
orbit, using this model, is very rapid and can achieve high
numerical precision. This model combines three axisymmetric potential-density forms to produce a model
of the matter distribution and its gravitational potential of the Milky Way.
The basic components of the Milky Way are the visible $\emph{disk}$ and $\emph{spheroid}$,
 and the invisible dark matter $\emph{halo}$.
For the bulge (spheroid) we adopt a Plummer sphere (Plummer 1911) in order to increase
the central mass of the galaxy. For the disk, we adopt Miyamoto-Nagai potential (Miyamoto \& Nagai 1975)
which is added in order to reproduce the scale length of the disk corresponding to $n = 0$. However, Evans \& Bowden (2014) and  Evans \& Williams (2014) presented new analytical families of axisymmetric
dark matter haloes, based on interesting modifications of the Miyamoto-Nagai potential, called Miyamoto-Nagai sequence.
And for the dark matter
halo we adopt a logarithmic (modified sphere)
potential, which produces a flat rotation curve at large radius, since for strongly flattened systems it is more
natural to work in cylindrical coordinates $R$, $z$, $\phi$ rather than spherical
$r$, $\theta$, $\phi$ (Binney et al. 1988;  Flynn et al. 1996).

\begin{equation}
\label{eq:(1)}
  \Phi=\Phi_{sph}+\Phi_{disk}+\Phi_{halo}\,
\end{equation}

where $\Phi_{sph}$, $\Phi_{halo}$ and $\Phi_{disk}$ define the spheroid, halo and
disk components respectively.

The first two components are described by the same law. For the disk component
we follow a Miyamoto-Nagai potential:

\begin{equation}
\label{eq:(2)}
  \Phi_{disk}(R,z)=-  \frac{GM_{disk}}{\sqrt{R^{2}+[A+(z^{2}+B^{2})^{1/2}]^{2}}}
\end{equation}

where it depends on two variables, namely \emph{R}, which denotes the radial distance perpendicular to the Galactic central axis, R$^{2}$= x$^{2}$ + y$^{2}$ and $z$, which denotes the vertical distance from the Galactic plane. The parameter $A$ is a measure of the radial scale-length of the disc
while the parameter $B$ is a measure of the disc thickness in the $z$ direction. Galactic discs are much larger in the radial than in
the vertical directions thus the values $A$ will be greater than those of the $B$ parameter in
our models.
The corresponding values for the disk component are $A=3.7$ ~kpc, $B=0.20$~kpc
 and $M_{disk}=8.01\times10^{10}$ ~M$_{\odot}$.

 Using Poisson's equation, $\nabla^{2}\phi$ = 4$\pi \rho G$, we can obtain the expression for the disc
density as

\begin{equation}
\label{eq:(3)}
\rho (R,z)= \left(\frac{B^{2}M}{4\pi}\right) \frac{A\emph{R}^{2}+(A+3\sqrt{z^{2}+B^{2}})(A+\sqrt{z^{2}+B^{2}})^{2}}{[\emph{R}^{2}+(A+\sqrt{z^{2}+B^{2}})^{2}]^{5/2}(\sqrt{z^{2}+B^{2}})^{3/2}}
\end{equation}

Note that when $A$ = 0 the potential reduces to a spherical potential on the galactic plane
($z$ = 0). The spheroid component of the Galactic gravitational potential is similar to the above:

\begin{equation}
\label{eq:(4)}
  \Phi_{sph}(R,z)=-  \frac{GM_{sph}}{\sqrt{R^{2}+[A+(z^{2}+B^{2})^{1/2}]^{2}}}.  \\
\end{equation}

The spheroid component, $A=0.0$ ~kpc, $B=0.28$ ~kpc and
$M_{sph}=1.12\times10^{10}$ ~M$_{\odot}$. Here we must point out, that this potential is not intended to represent a black hole nor any other compact object, but a dense and massive bulge. Therefore, we do not include any relativistic effects (Jung \& Zotos 2015).

Regarding the halo component of the Galactic
gravitational potential, is given by the following
equation:
\begin{equation}
\begin{aligned}
 \label{eq:(3)}
 \Phi\rm _{halo}=\frac{GM\rm_{halo}}{r_{c}}
  \left[\frac{1}{2}\ln\left(1+\frac{R^{2}+z^{2}}{r_{c}^{2}}\right) +
  \right.\\
\left.\rm \frac{\rm r_{c}}{\rm \sqrt{\rm R^{2}+z^{2}}}\arctan
\left(\rm \frac{\rm \rm \sqrt{\rm R^{2}+z^{2}}}{\rm
r_{c}}\right)\right]
\end{aligned}
\end{equation}
where $\rm r_{\rm{halo}}=12$~kpc and $\rm M_{\rm{halo}} = 5.0 \times
10^{10}$~$\rm M_{\odot}$.

While in a spherically symmetric potential the $x$, $y$ and $z$ components of
the angular momentum and the total angular momentum are conserved, out potential only
admits two conserved quantities, the total energy $E$ and the $z$-component of
the angular momentum vector. Therefore, the motion of the objects are in the plane and perpendicular
to the vector of the total angular momentum. We assume that the rotation curve is composed of a
spheriod, disk, averaged circular velocity and triaxial halo components in the Galaxy (see Fig. \ref{rotation-curve}), and is calculated at any radius by
$v^{2}_{c}(r) = r\frac{d\Phi}{dr}$. It is clearly seen from the same figure that each contribution prevails in
different distances form the galacto-centric $R$. In particular,
at small distances when $R \geq 3$ kpc, the contribution from
the spheriod dominates, while at distances of $3 < R < 19$ kpc the disk contribution is the dominant
factor. On the other hand, at large enough galacto-centric
distances, $R > 19$ kpc, we see that the contribution
from the triaxial halo prevails, thus forcing the rotation curve
to remain at with increasing distance from the center (Zotos 2014).

\subsection{NS Initial Velocities}
\label{ourselection}

The initial velocities of the NSs are calculated as the vector addition
of three different velocities: (1) a Maxwellian distribution, (2) a constant kick,
and (3) the circular rotation velocities at the birth place.
The details on their calculation can be found in Arzoumanian et al. (2002) and
Hobbs et al. (2005) and are summarised here.

Concerning the density distribution of the total number of NSs which
we used in this simulation, it is about $\sim$ $1 \times10^{7}$, and it
distributes uniformly in the age ~  $\geq 1 $~Gyr. This is equivalent to a simulating
birthrate of one NS per century
(Kulkarni \& van Kerkwijk 1998; Lyne \& smith 2007).
However, the precise estimates of both the number and lifetime of the NS population are
hard to obtain especially at larger distances
(see, e.g Keane \& Kramer 2008; Ofek 2009)
because they may have been heavily biased by a number of
observational selection effects (Lorimer 2008).


Regarding the kicks simulation, the kick velocity imparted to an NS at birth is one of the major
problems in the theory of stellar evolution. However, the physical
mechanism that causes this kick is presently unknown and a number of physical models
have been described and evaluated
(see, e.g. Kalogera 1996; Wang \& Han 2010; Meng et al. 2009). But it is presumably the
result of some asymmetry in the core collapse or
subsequent SNe explosion (see, e.g., Pfahl et al. 2002; Podsiadlowski et al. 2004).
However, the distribution of the kick amplitudes is usually obtained from the analysis
of radio pulsar proper motions (Hobbs et al. 2005).

The short lived sudden kick in the phase space has a kick direction uncorrelated
with the orientation of the plane and the system velocity that the object acquires
because the explosion is also uncorrelated with the rotation in the Galaxy.
Therefore, we follow Hansen \& Phinney 1997) and find that the distribution is consistent
with Maxwellian distribution of kick velocity. Finally, regarding the third element used in the calculation of the
initial condition, we choose the initial circular rotation velocity of
the NS before a kick. A plot of averaged circular velocity  for every parts of our galactic potential model presented is in Fig. \ref{rotation-curve}. It rises linearly up to a maximum value and then they become constant for larger radii.

A distinct approach to the analysis of location and velocity of old NSs is based on the Monte Carlo simulation of the evolution of a simulated sample with different initial parameters,  and then their
orbits are numerically integrated. We would utilize the method of the Poincar\'{e} sections to explore their
motions. This
has been studied in detail in Papers I \& II). Once we have selected the above, we can chose a set of initial conditions.
We will consider isolated NSs (excluding the MSPs and GCs), being of ages older than
$10^9$ yrs and with large radial expansions, $10 > \emph{R} > 25$ \emph{Kpc}.
The radial distribution has an exponential scale length of $\lambda e^{-\lambda |z|}$, where $\lambda =1/0.07 ~kpc^{-1}$. Here we assumed a maximum birth height off the plane, $\emph{z}_{max}$ of $150$ \emph{pc}.

\section{Numerical Results}

\subsection{Selection of initial conditions}

We deal in this section with the analysis of the dynamical evolution of
a representative set of NSs' Orbits in the Galaxy calculated
following the previously described distribution of velocities.
We aim to obtain a set of different orbits representing NSs at given distances from the axis.
Our method follows Arzoumanian (2002). We start by choosing a randomly selected initial position, with boundaries
located at the Galacto-centric distance in the $r \leq 25 kpc$.

The initial velocity components of the NSs are also obtained from a random sample,
following the vector addition of the three different velocities described in the previous section: the Maxwellian
distribution, the constant kick and the circular motion velocities at the birth place.
The direction of the initial velocity vector is chosen randomly within the geometric shape
of the potential.
We start from a vector based on the the circular velocity. Then we add a
random kick, in the form of a local perturbation, with modulus following the Maxwellian
distribution (Hansen \& Phinney 1997) and direction to
the radial direction.
Finally, we add the third vector. This is a velocity vector with a modulus fulfilling
a Maxwellian distribution with $\sigma_v$ = 265 km s$^{-1}$ (Hobbs et al. 2005; Story \& Gonthier 2007) and $\sigma_v$ =190 km s$^{-1}$ (Hansen \& Phinney 1997). The direction is given by selecting randomly a direction
specified by choosing two angles, $0 < \Phi < 360 $ and
 $ - 90 < \theta < 90 $ (Kiel \& Hurly 2009).

 The equations above were solved numerically in three directions using the
5th order Runge-Kutta method with adaptive control step sizes (Press et al. 1992). The
initial step size is $\emph{dt} = 10^{-4}$ Myr, and calculations of position
and velocity in 3-D then continues until 0.1 Myr with adjusted step sizes according to the required accuracy.
The data are recorded every 0.1 Myr and the total energy
$E_{tot} = \frac{1}{2}(v^{2}_{x} + v^{2}_{y} + v^{2}_{z} ) + \Phi(\emph{R}, \emph{z})$ are checked. The position
and velocity then are used for input of the next 0.1 Myr.
The procedure used here enabled us to control and achieve required
levels of accuracy by using the energy integral.

The Monte Carlo simulations were run and the results are given here.
One hundred initial conditions were obtained.
Different families are identified and we have selected O1, O2, O3.
These orbits are listed in Table 1. A periodic orbit is found when the initial and final coordinates coincide
with an accuracy at least $10^{-15}$.
This is a representative set because they show the typical behavior
of the whole distribution, with very large Galactic radial expansion and they are rosette-like
type orbits. In addition, the orbits lie essentially in the plane of the long and intermediate axes and so reinforces the shape of
the potential to some extent. The most general trajectory for an object in this
potential, which allowed us to clarify properties
on the structure and configuration of the invariant tori that we encounter in the vicinity of the periodic orbits.
The exact initial conditions for the
periodic orbit are calculated and listed in Table~1.
The trajectories and 2-D projections corresponding to the integration of every
initial condition up to a timescale of $10^{8}$ \emph{yr} are presented in Figs. 2-5.

\begin{table*}[ht]

\label{initialconds}
\begin{tabular}{lcccc}
\hline \hline \noalign{\smallskip}
Orbit Label & Initial position $(x_0,y_0,x_0)$ &  Initial position $(r_0,\theta_0,\Phi_0)$ & Initial velocity ($v_{x0},v_{y0},v_{z0})$ & $\lvert v \rvert$ \\
\hline \noalign{\smallskip}
O1 & $(-1.11, -9.2, 0.008)$ & $(9.267, 0.049, -96.87)$ & $(172.4, -53.0, -121.2)$ & 217.302 \\
\noalign{\smallskip}
O2 & $(-2.14, -4.0, 0.1)$ & $(4.538, 1.26, -118.14)$ &  $(15.8, -161.8, -137.4)$ & 212.856 \\
\noalign{\smallskip}
O3 & $(6.88, -6.78, 0.2)$ & $(9.661, 1.18, -44.58)$ &  $(48.5, -44.7, -123.7)$& 140.185\\
\hline
\end{tabular}
 \caption
 {Selected Representative orbits, $\theta$ and $\Phi$ in degrees, $\lvert v \rvert$ in km s$^{-1}$ }
\end{table*}

\begin{table*}[ht]
\label{initialconds}
\begin{tabular}{lcc}
\hline \hline \noalign{\smallskip}
Orbit Label & $\lambda (\phi = 0) $ & $\lambda (\phi = 90) $ \\
\hline \noalign{\smallskip}
O1 & 0.00015160387 & 0.00015288824 \\
\noalign{\smallskip}
O2 & 4.5184949  & 4.3155085  \\
\noalign{\smallskip}
O3 & 1.9746514 & 2.0018652 \\
\hline
\end{tabular}
 \caption{Maximum Asymptotic Lyapunov exponents for the selected Representative orbits,
 two different orientations of dark halo.}
\end{table*}
\newpage

As we can see from Fig. 2, the different initial heights have an influence
on z-direction due to the perturbations and will naturally cause different
trajectories, while if the objects are located at
different R, trajectories can be less different and may prone to instability through the gravitational perturbation.

It is noteworthy to mention here that if the sun received a kick velocity ($~$50 km/s)
in the Galactic plane (Repetto et al. 2012a), the typical orbit (considered as a circular
 in the Galactic potential) would turn into a rosette orbit.

\subsection{Dynamical Analysis}

\subsubsection{Poincar\'{e} sections}

In Poincar\'{e} sections, periodic orbits appear in the
surface of section as a finite set of points. The amount of points depends
from the multiplicity of the periodic orbit.
This method has been extensively
applied to 2-D Hamiltonians in a 2-dimensional plane (Lichtenberg \& Lieberman 1992; Wei et al. 2010a; Taani et al. 2012b). While in 3-D systems, we have to project the Poincar\'{e}
surface to spaces with lower dimensions (Contopoulos 2004).

By setting $\emph{y}$, $v_{y0}$ equal to zero at t = 0 (remaining zero at all
times) in the Hamiltonian equation, the 3-D Poincar\'{e} sections can be found in Fig.~3.
The points in these figures show the 3-D section. We generate 2500 consequents
of the orbits in 3-D ($x$, v$_{x}$, $z$).
We find that there are sometimes several
orbits from the same family plotted very close to each other. This happens
when the sequence of orbits within the family reverses its
progression in the $x$ - $z$ plane towards a given direction
and very close to the current space position, causing an accumulation
of orbits near this position with different local
velocities. The stable periodic orbits are surrounded by
quasi-periodic orbits that bound to the surface of 3-D
tori in a 3-D Hamiltonian system. These quasi-periodic orbits are represented in the
surface of section by 2-D tori (Manos et al. 2012).

\begin{figure*} 
\begin{center}
\begin{tabular}{cc}
\includegraphics[width=6cm]{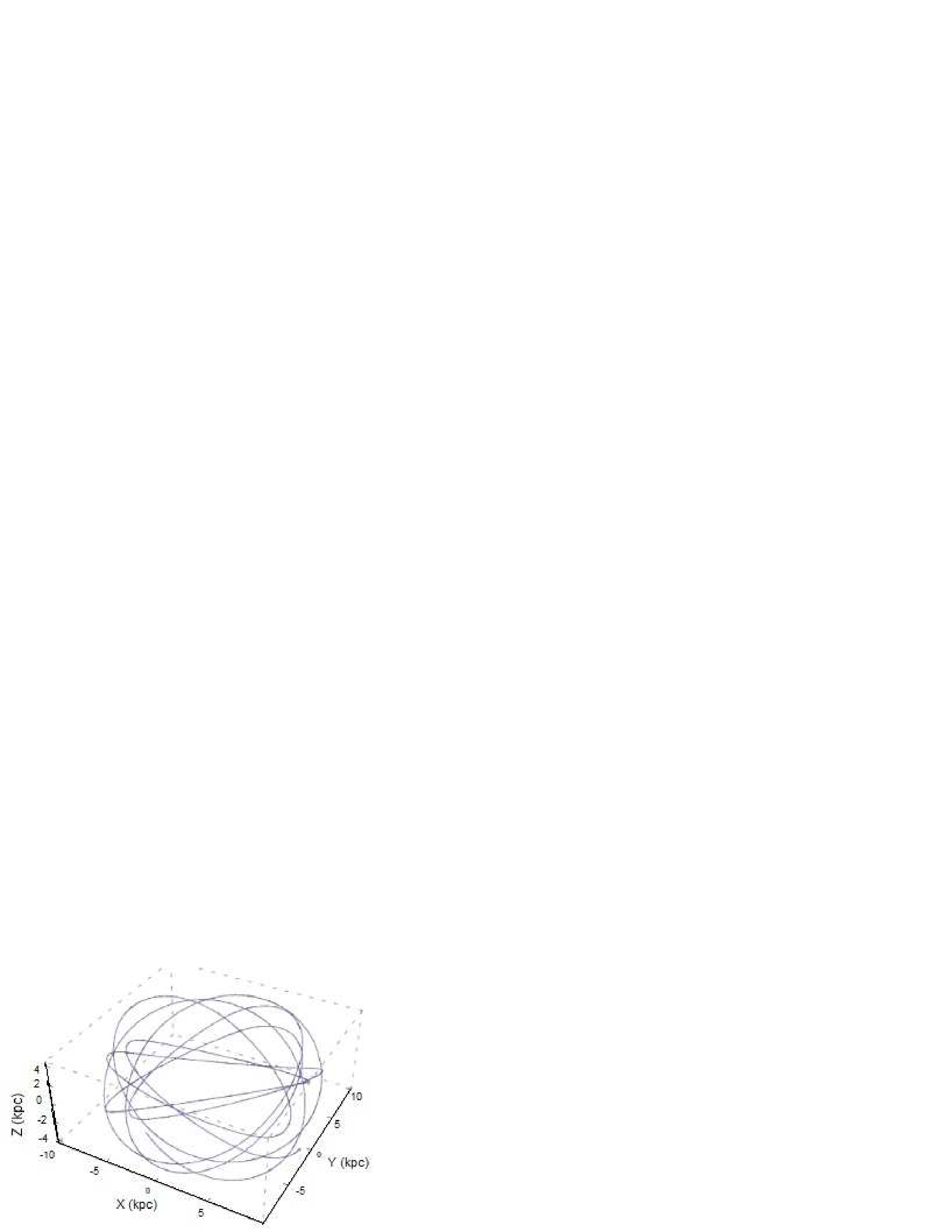}
\includegraphics[width=6cm]{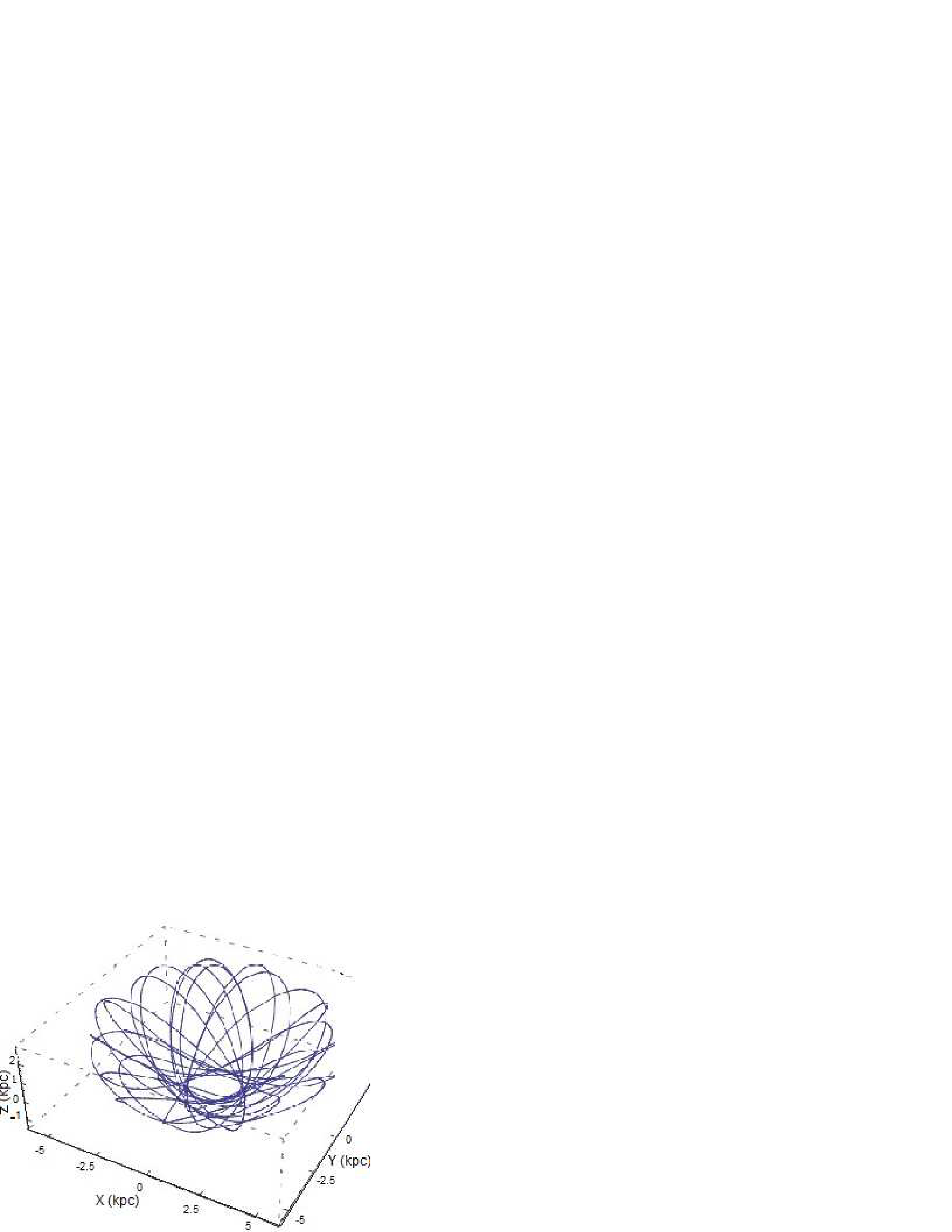} \\
\includegraphics[width=6cm]{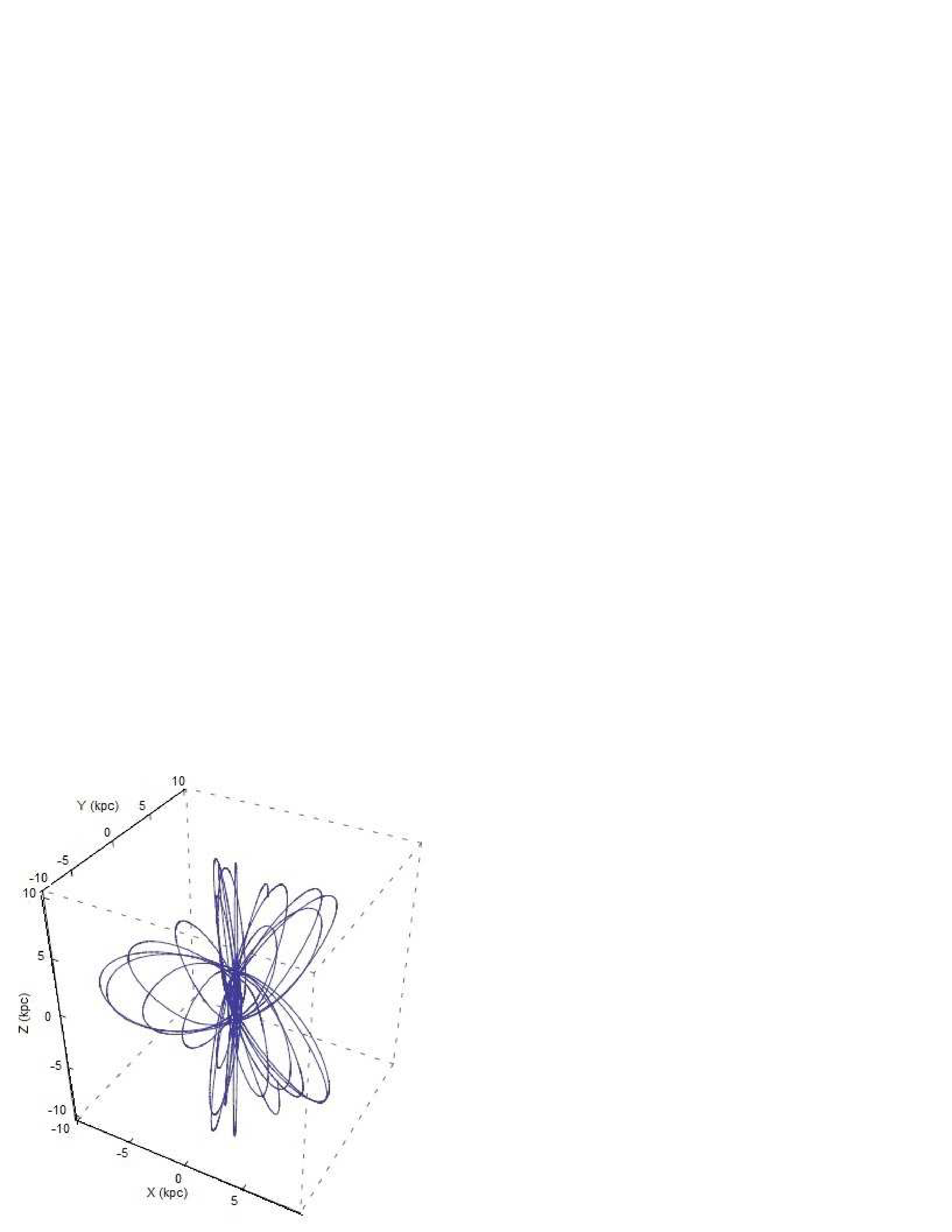}
\label{fig1}
\end{tabular}
\end{center}
 \caption{Trajectories corresponding to the different initial conditions of
 Table~\ref{initialconds}.  These conditions are given
in enlarged scale, in order to better view the corresponding
morphology. The upper left panel corresponds to orbit O1.
 The upper right panel corresponds to orbit O2. \textbf{The trajectory is recognized as a precessing
banana orbit (See Evans \& Bowden 2014)}. The bottom left panel corresponds to orbit O3.}
\end{figure*}

\begin{figure*} 
 \begin{center}
 \begin{tabular}{cc}
\includegraphics[width=6cm]{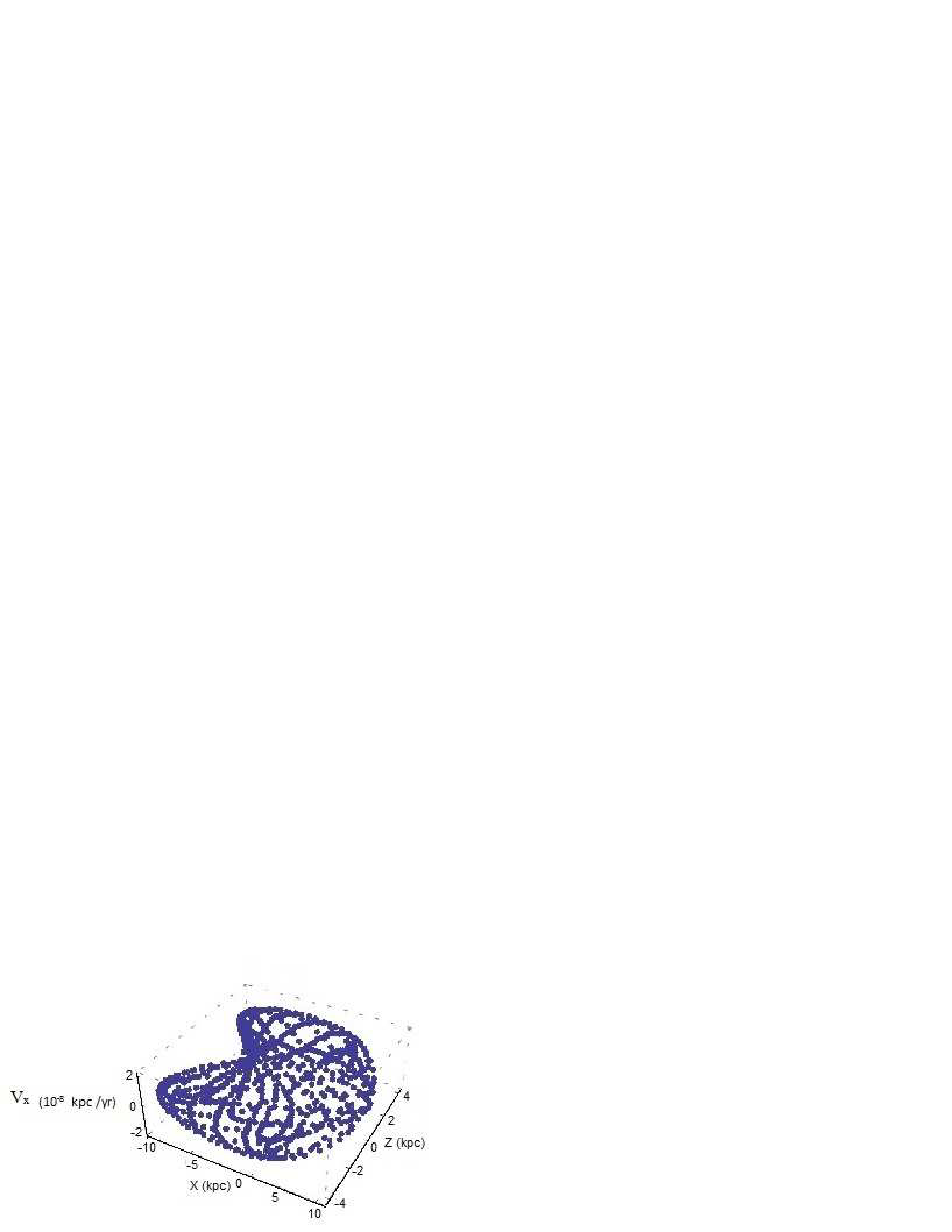} &
\includegraphics[width=6cm]{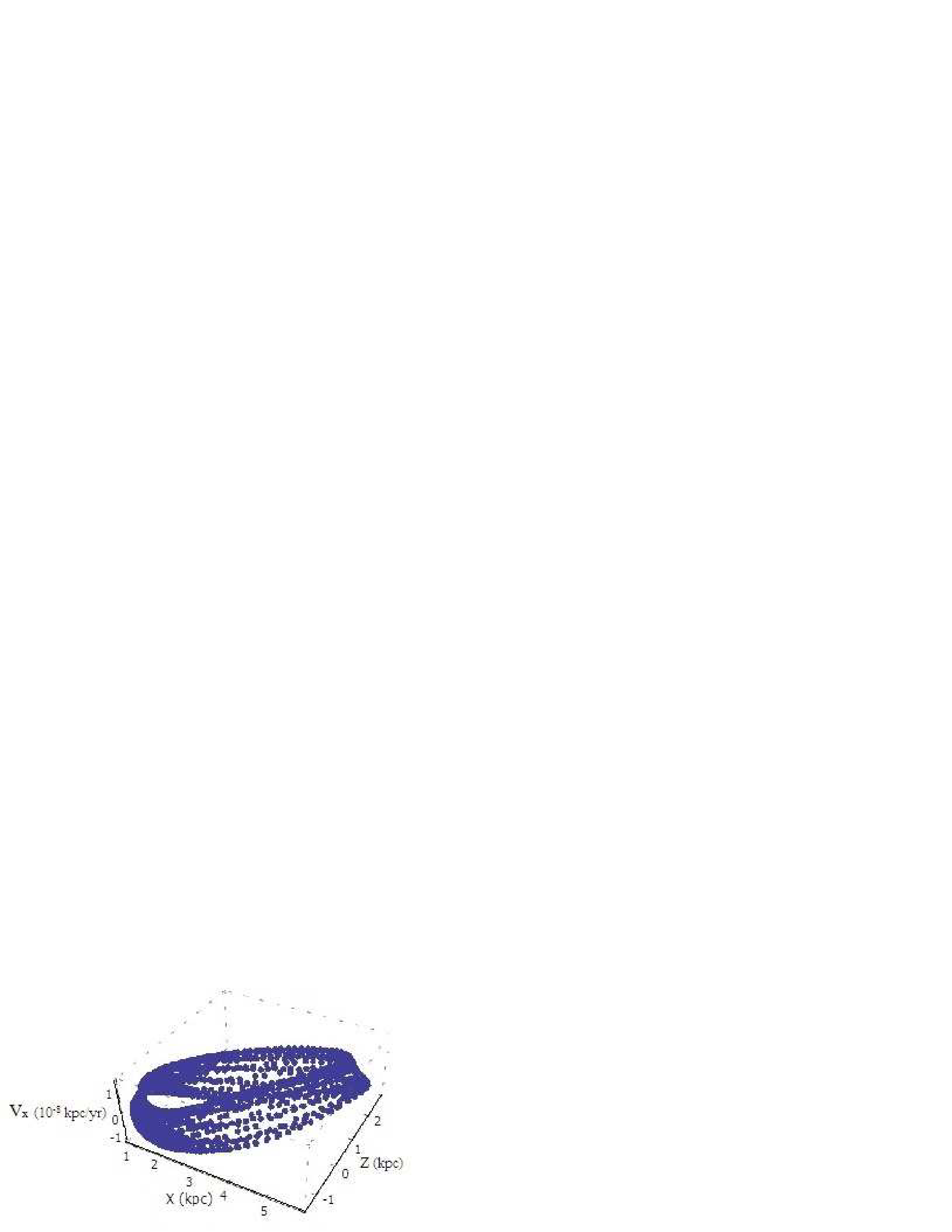} \\
\includegraphics[width=6cm]{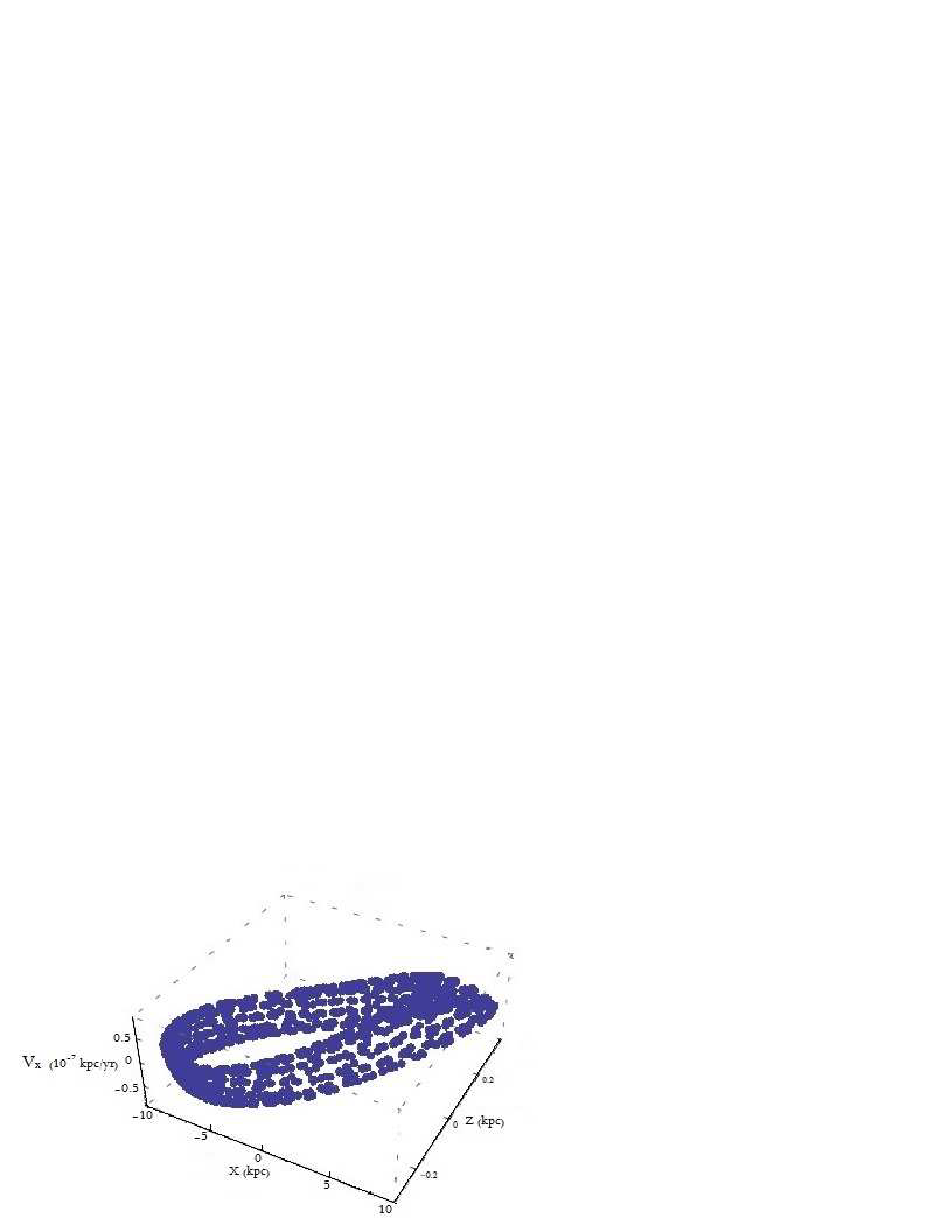} &
\label{fig3}
\end{tabular}
  \end{center}
  \caption{ The 3-D Poincar\'{e} section $x > 0$, at y = 0 of the 3-D trajectories
  corresponding to the selected initial conditions.
  The upper left panel corresponds to orbit O1.
 The upper right panel corresponds to orbit O2.
 The bottom left panel corresponds to orbit O3.}
\end{figure*}


Aiming to gain a better picture, Fig. 4 shows the 2-D projection in $x-z$ plane,
meanwhile Fig.5 shows the 2-D projection in $x-v_{x}$.
In the first figure, the intersection points distribute in some regular lines
 on the Poincar\'{e} hyper-surface section,  while the $x$ - v$_{x}$ projections are tend to
have enclosed curves. This behavior could indicate a regular motion and an quasi-periodic
 orbit. All the points along the orbits form ensembles of invariant subset of phase-space.
The ensemble of old NSs along given orbits for invariant blocks of the
  galaxy, since the galaxies are seen as built not of stars, but of orbits (Guarinos 1992).

\begin{figure*} 
 \begin{center}
 \begin{tabular}{cc}

\includegraphics[width=6cm]{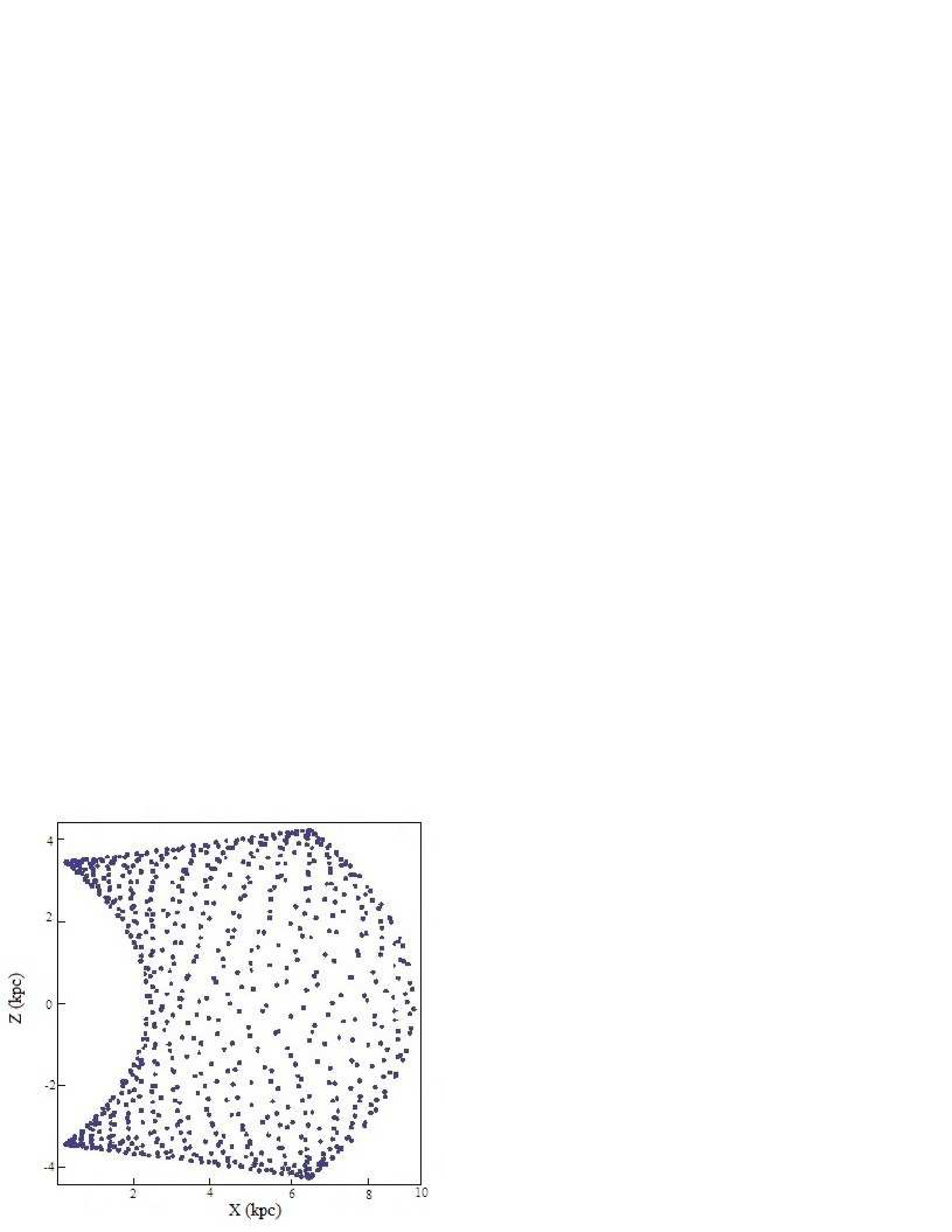} &
\includegraphics[width=6cm]{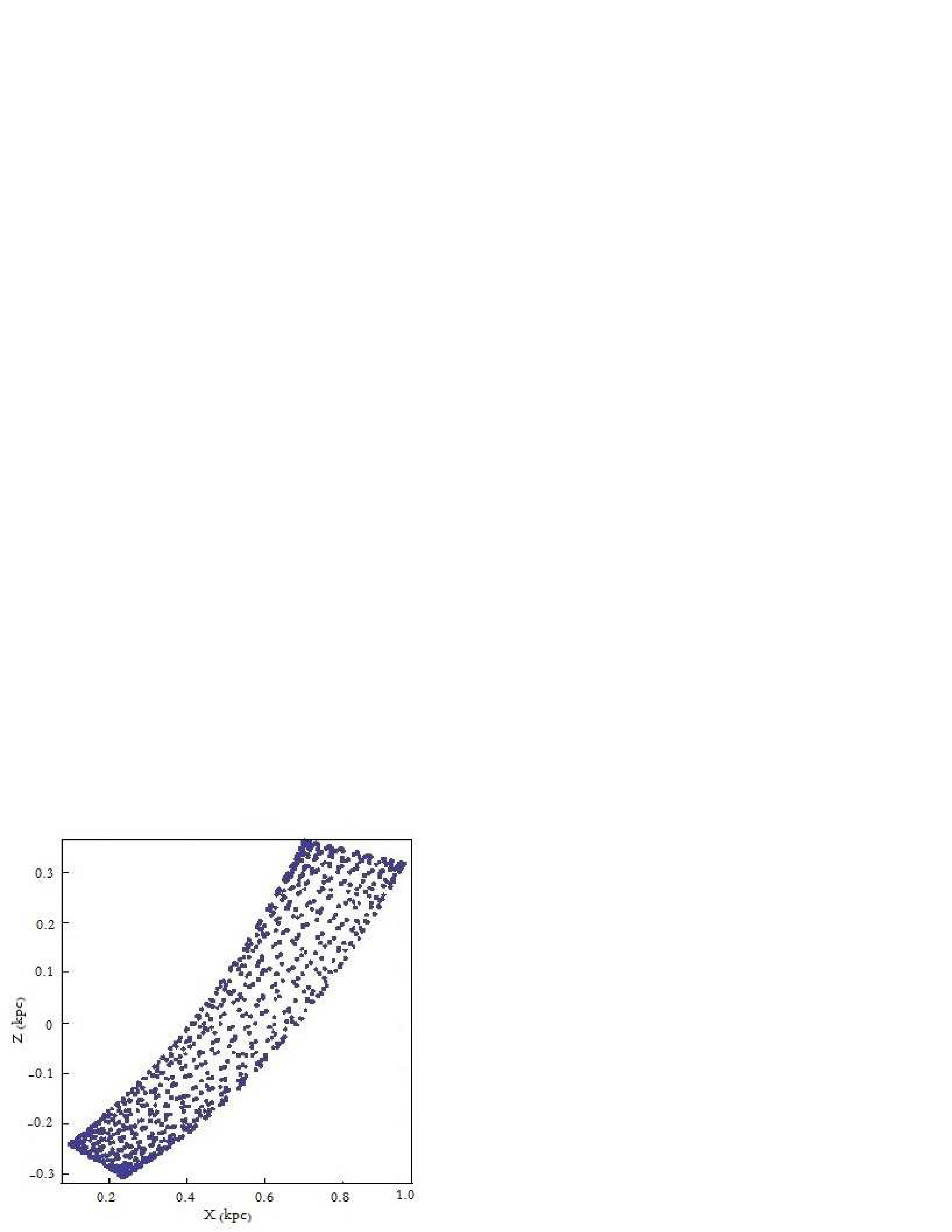} \\
\includegraphics[width=6cm]{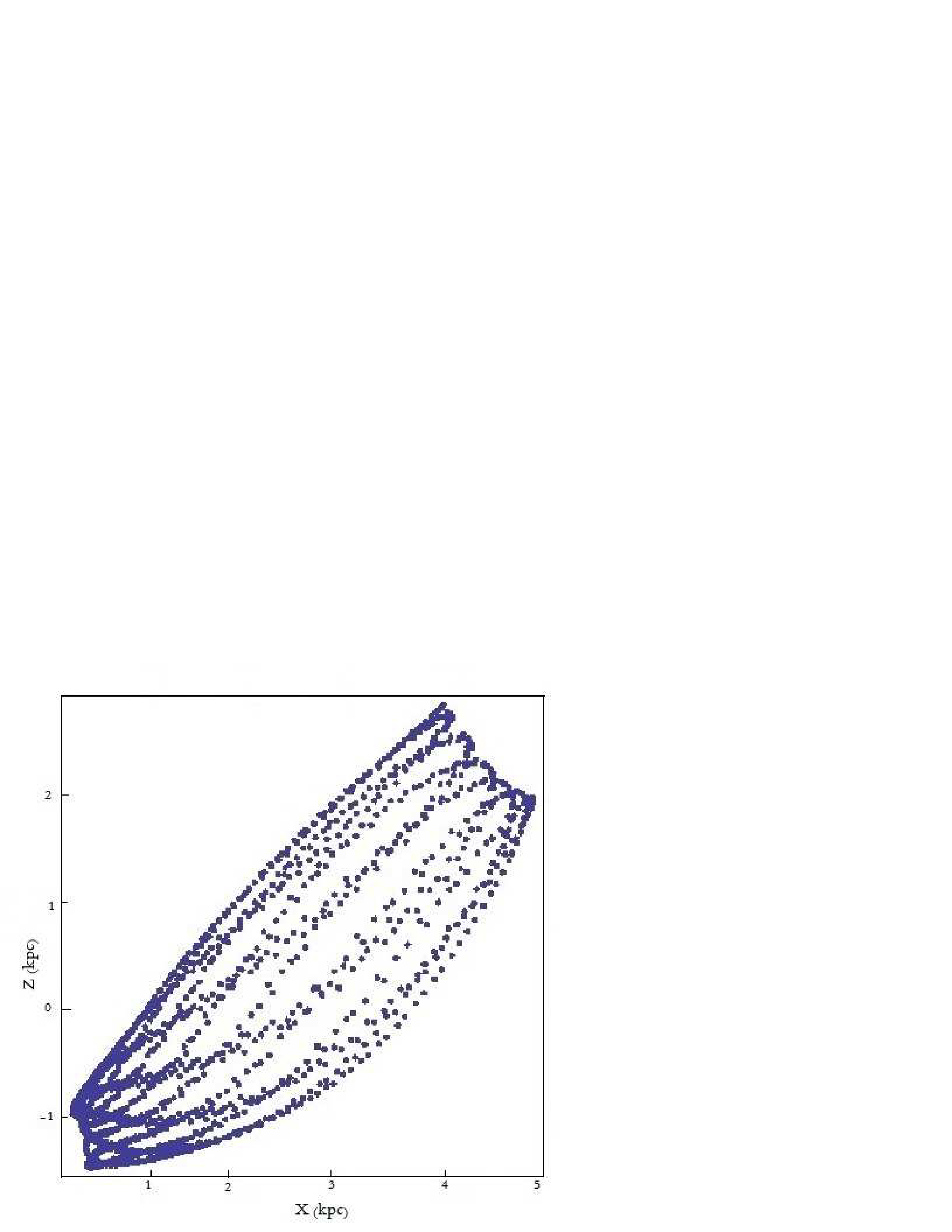} &
\label{fig4:projection}
\end{tabular}
  \end{center}
  \caption{ The projection on $x$ - $z$ plane of the 3-D Poincar\'{e} section.
  The upper left panel corresponds to orbit O1.
 The upper right panel corresponds to orbit O2.
 The bottom left panel corresponds to orbit O3.}
\end{figure*}

\begin{figure*} 
 \begin{center}
 \begin{tabular}{cc}
\includegraphics[width=7cm]{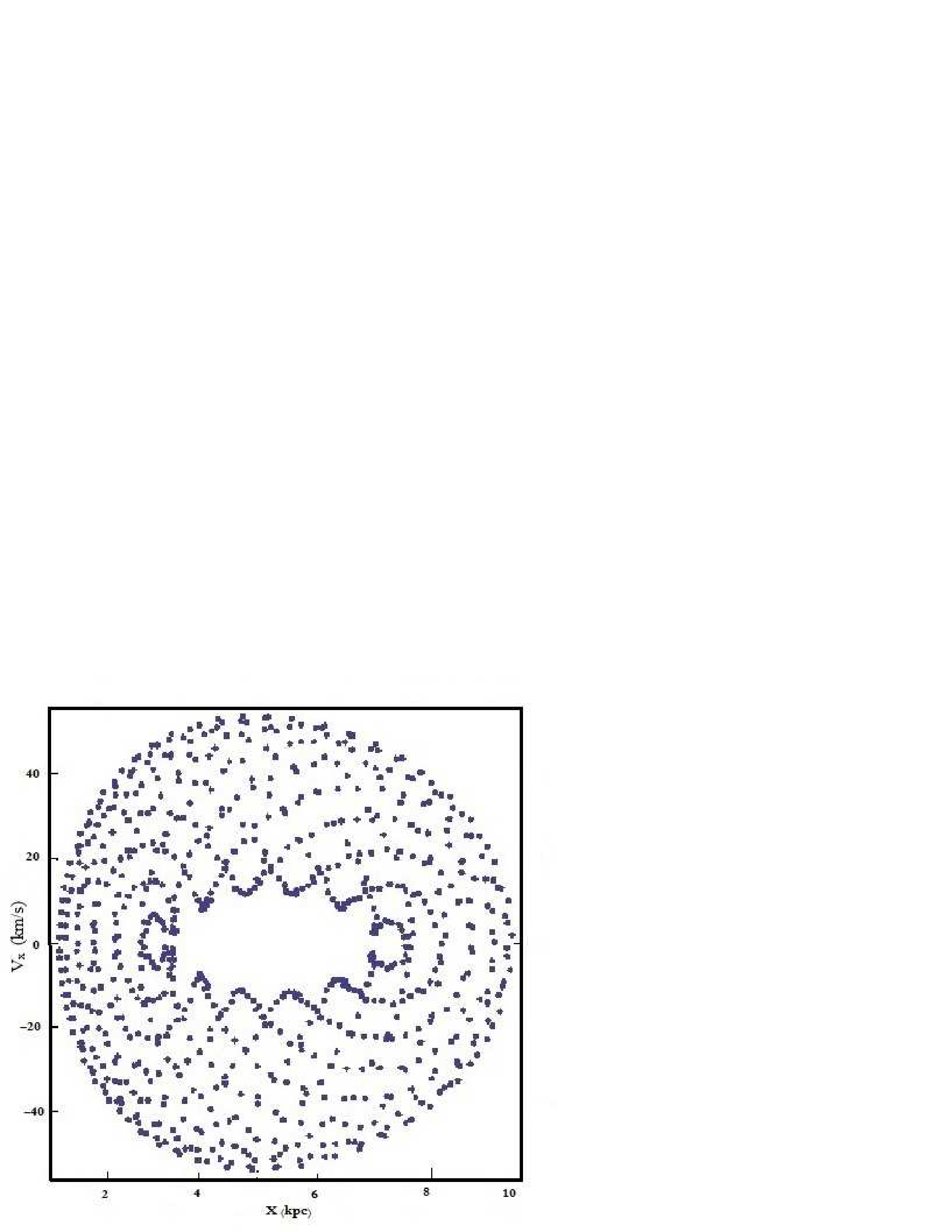} &
\includegraphics[width=7cm]{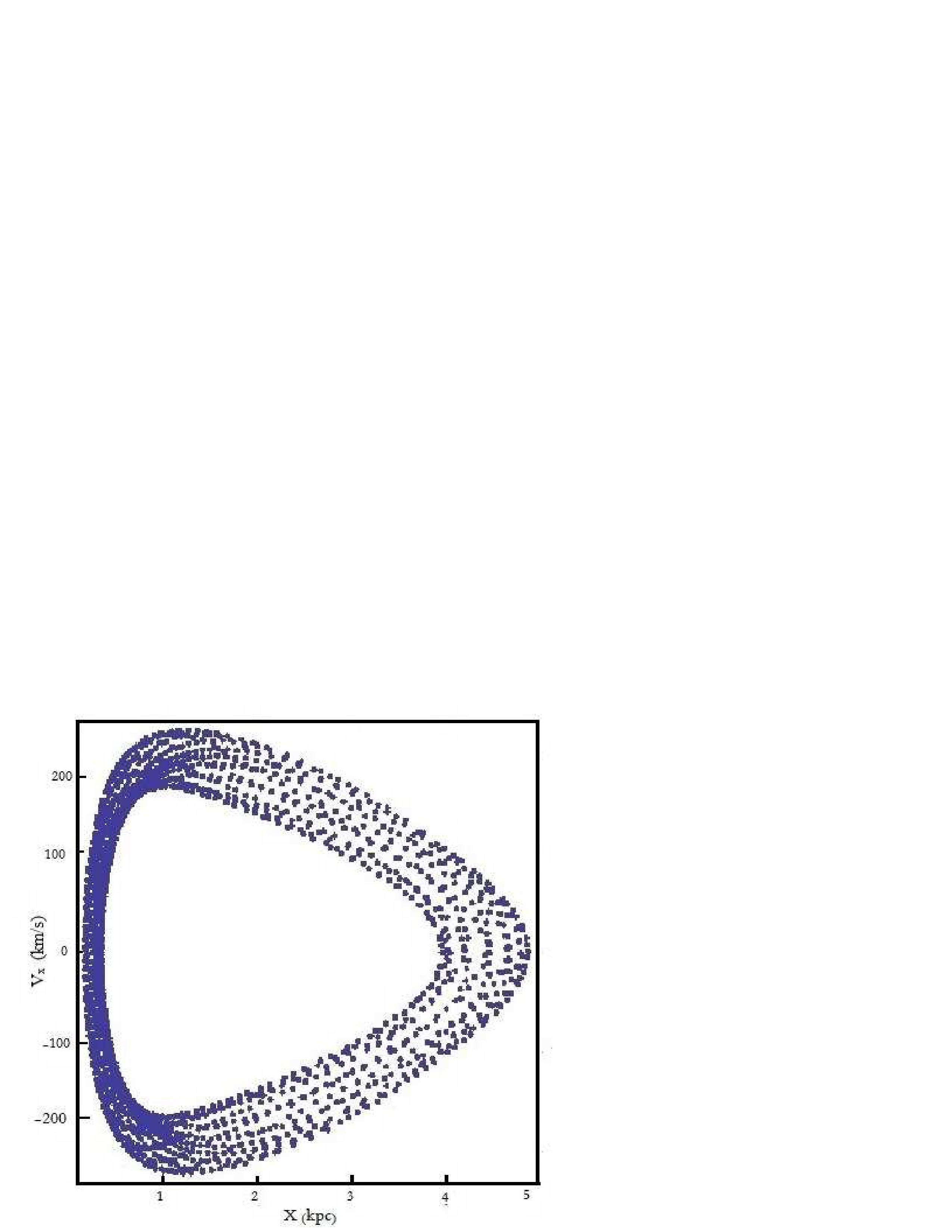} \\
\includegraphics[width=7cm]{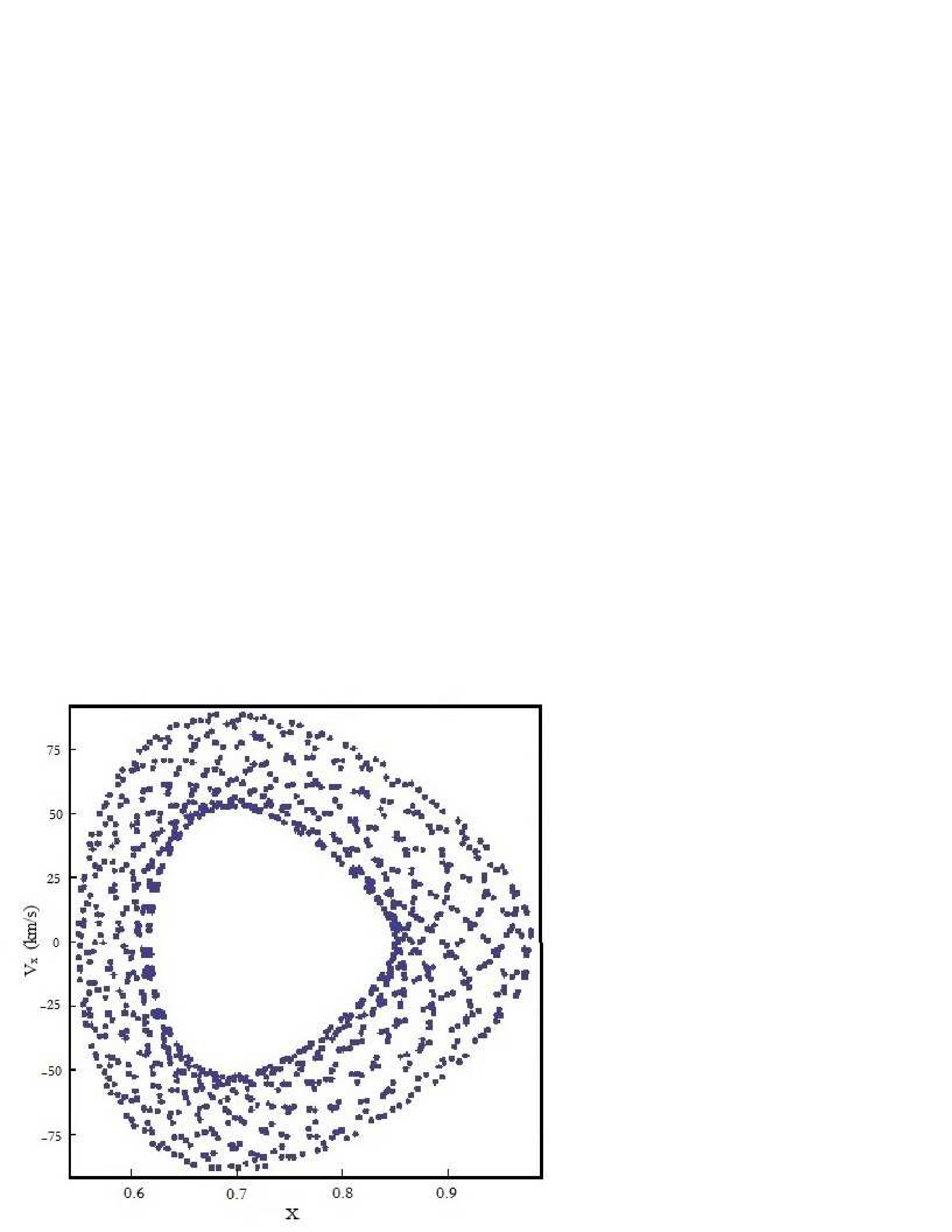} &
\label{fig5}
\end{tabular}
  \end{center}
  \caption{ The projection on $x$ - $v_{x}$ plane of the
3-D Poincar\'{e} section.
 The upper left panel corresponds to orbit O1.
 The upper right panel corresponds to orbit O2.
 The bottom left panel corresponds to orbit O3.}
\end{figure*}

\begin{figure*}

\begin{center}
\begin{tabular}{cc}
\includegraphics[width = 8cm]{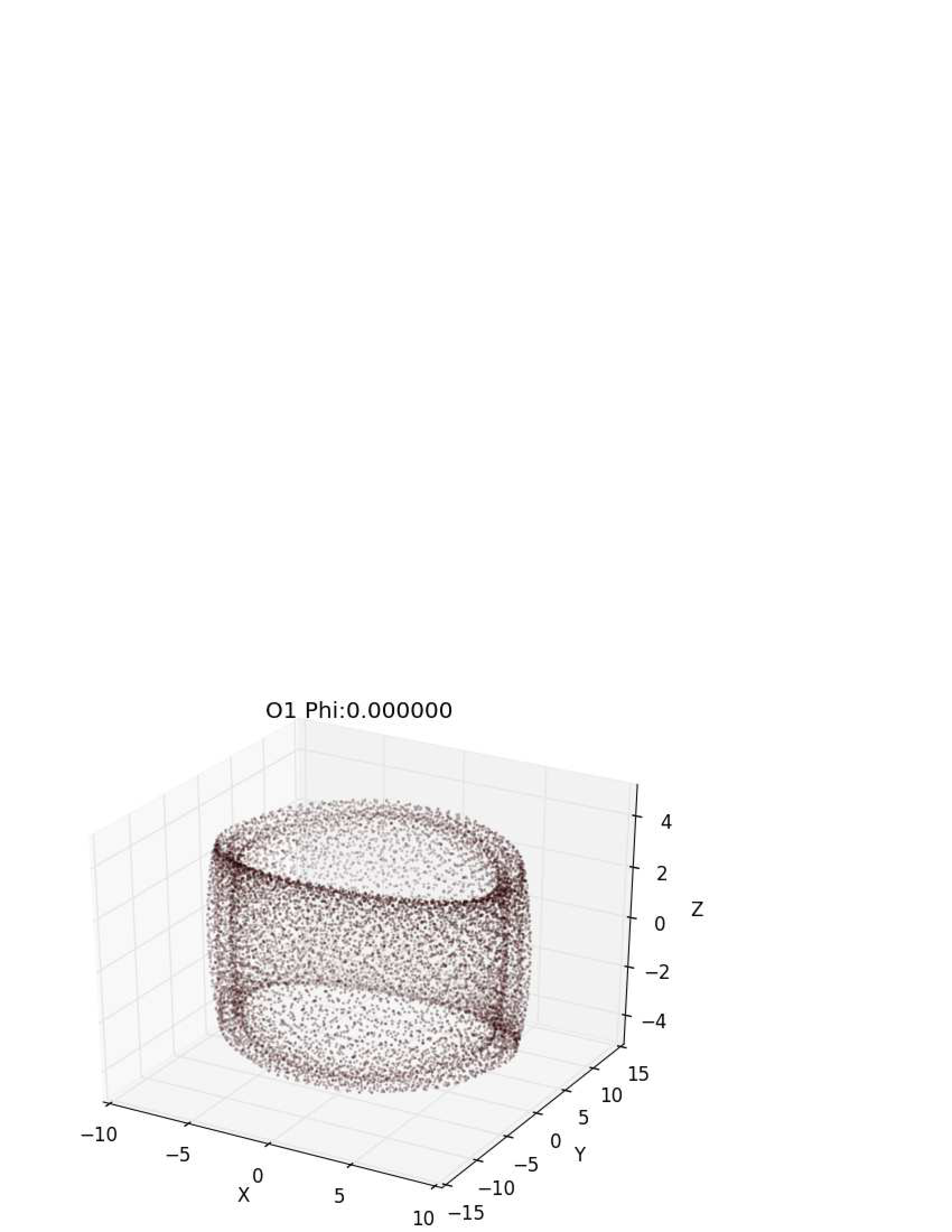} &
\includegraphics[width = 8cm]{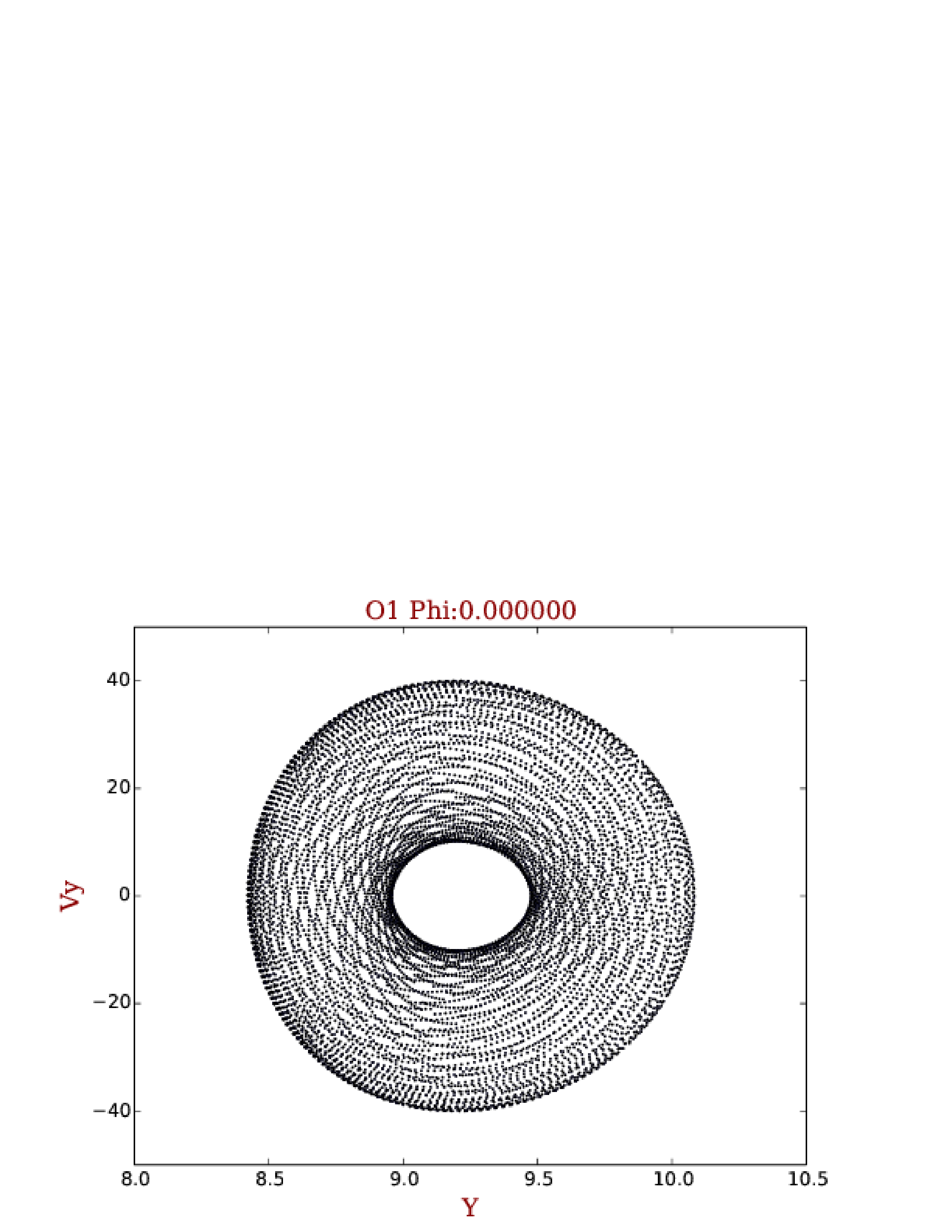}
\\
\includegraphics[width = 8cm]{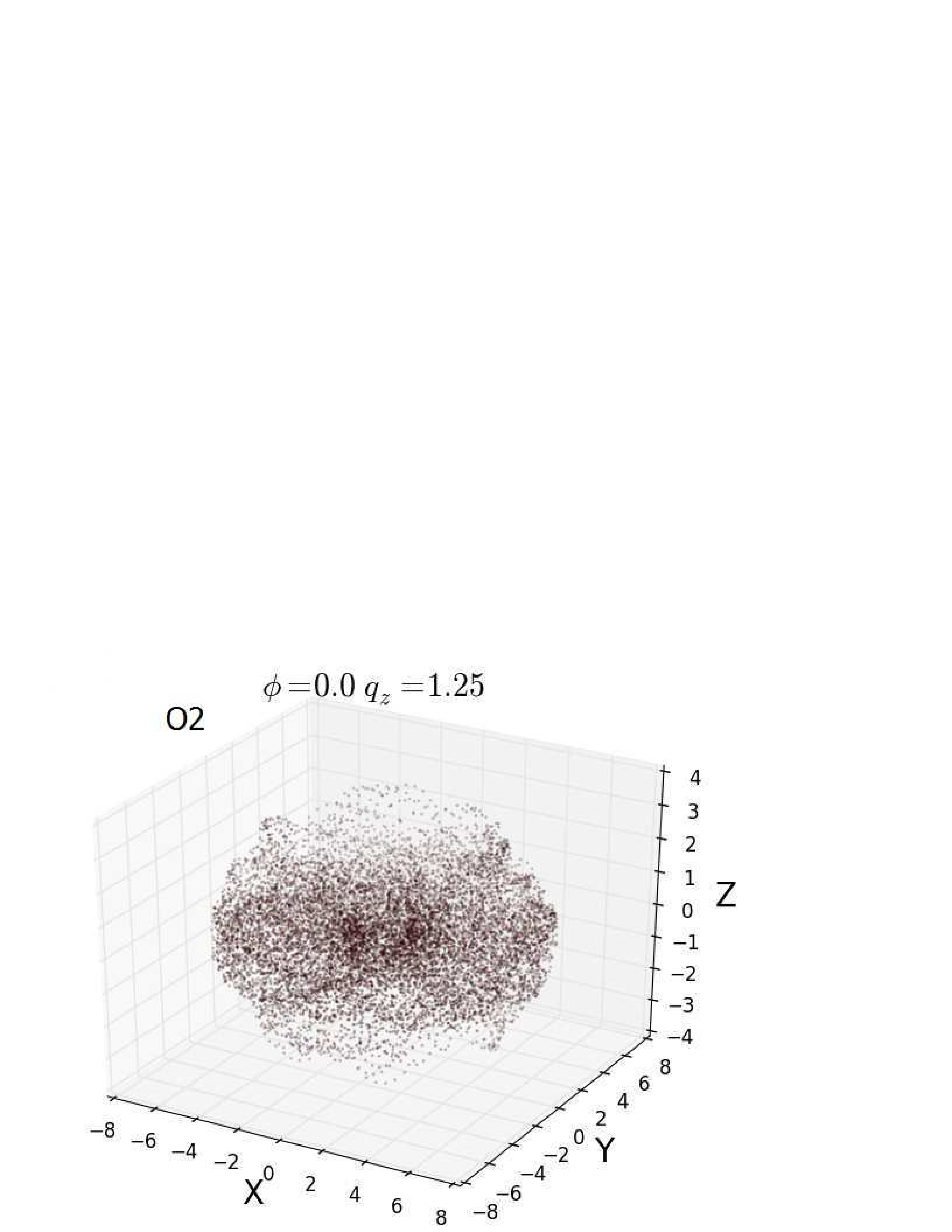} &
\includegraphics[width = 8cm]{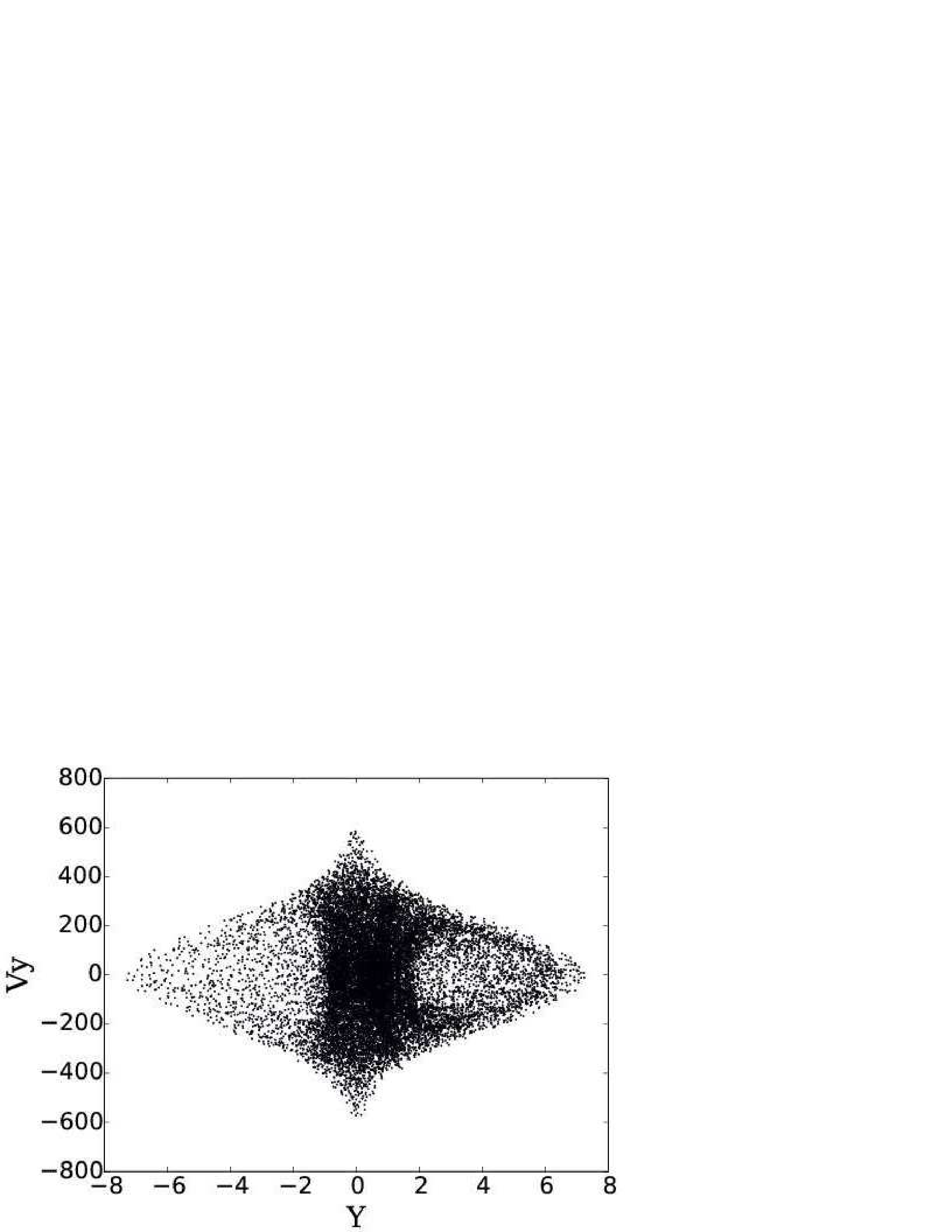}
\\
\includegraphics[width = 8cm]{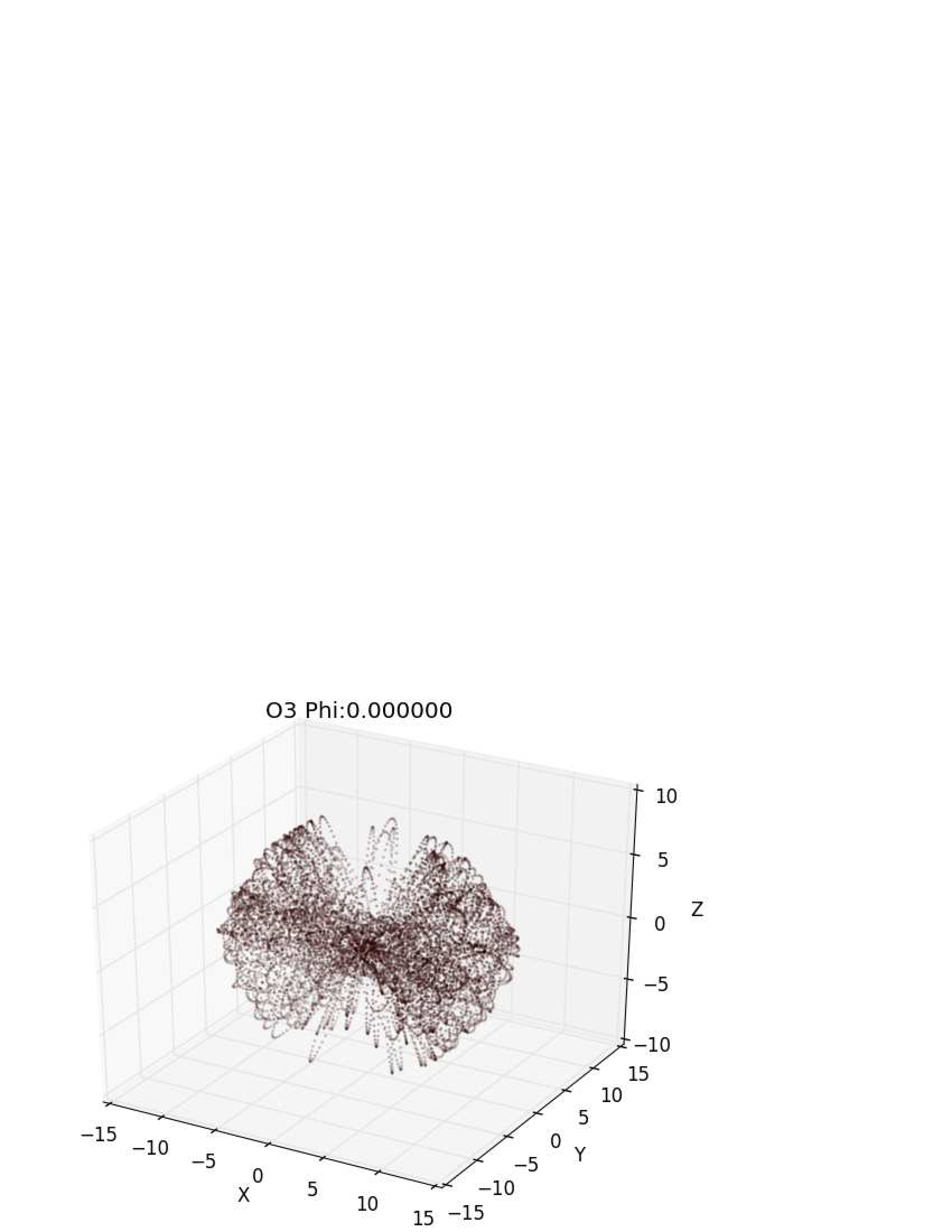} &
\includegraphics[width = 8cm]{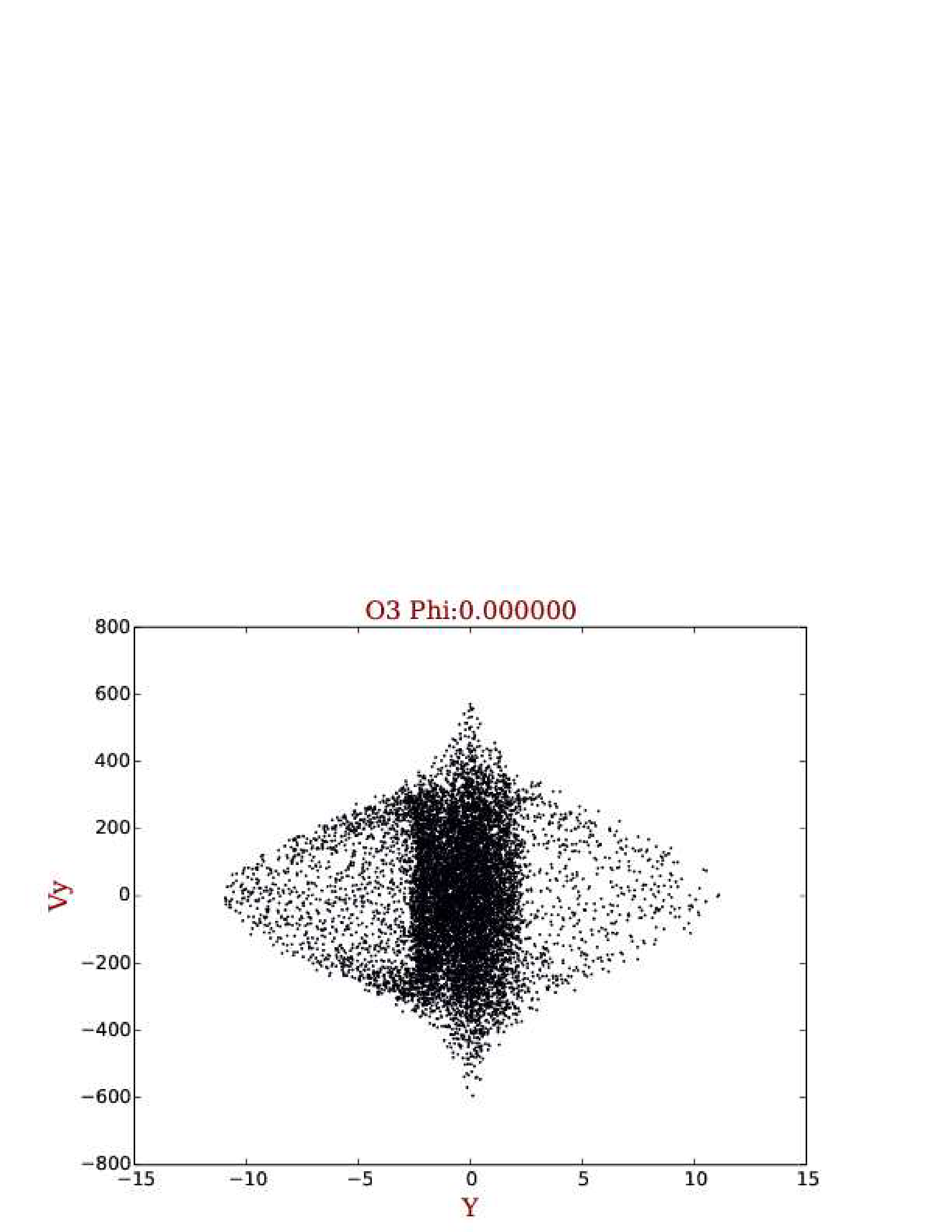}
\\
\end{tabular}
\end{center}

\caption{
Physical trajectories, and the corresponding
Poincar\'{e} sections $y - v_y$, with plane $x=0$, and $v_x > 0$,
for the selected Representative orbits, for a dark halo orientation
of $\phi = 0$.
}
\label{fig2}
\end{figure*}

\begin{figure*}

\begin{center}
\begin{tabular}{cc}
\includegraphics[width = 8cm]{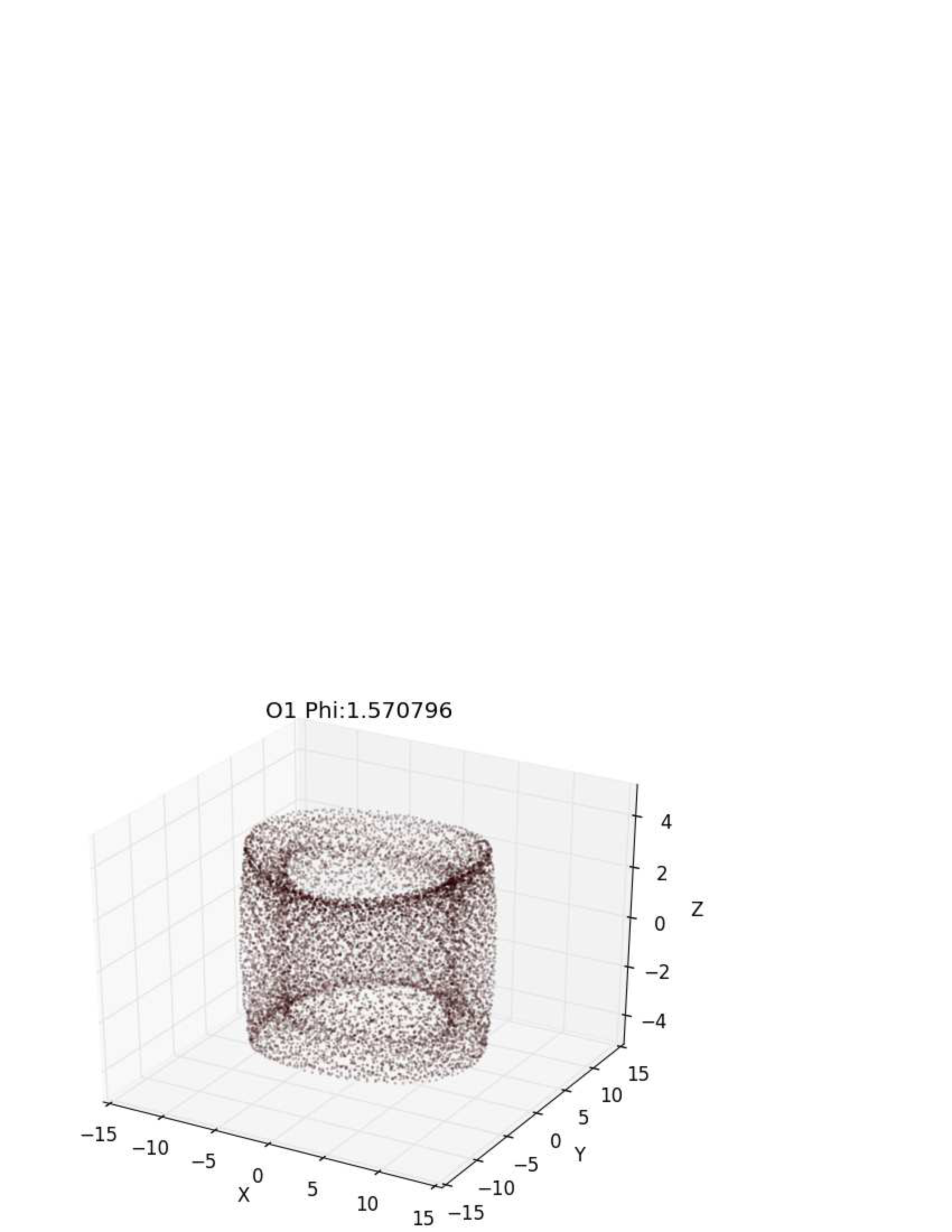} &
\includegraphics[width = 8cm]{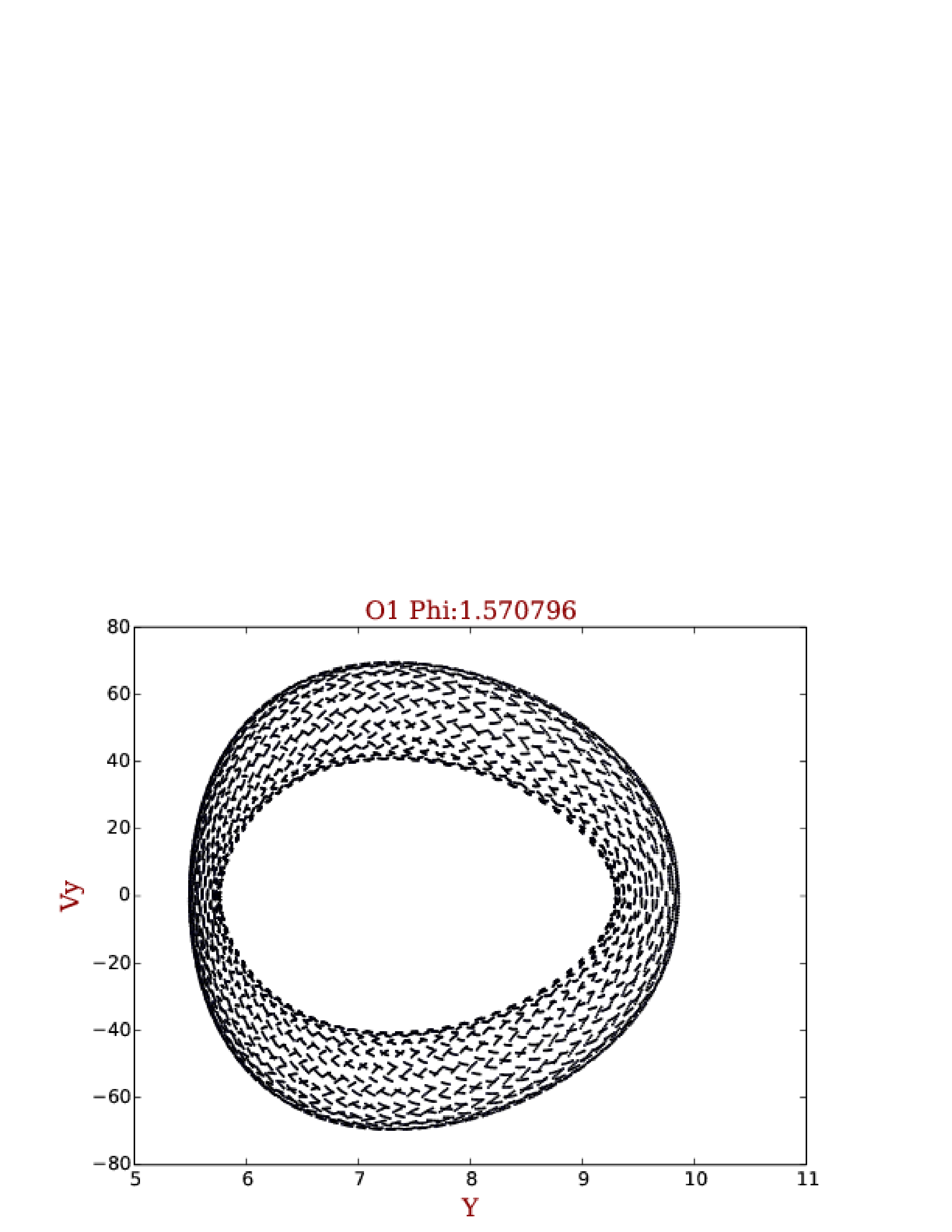}
\\
\includegraphics[width = 8cm]{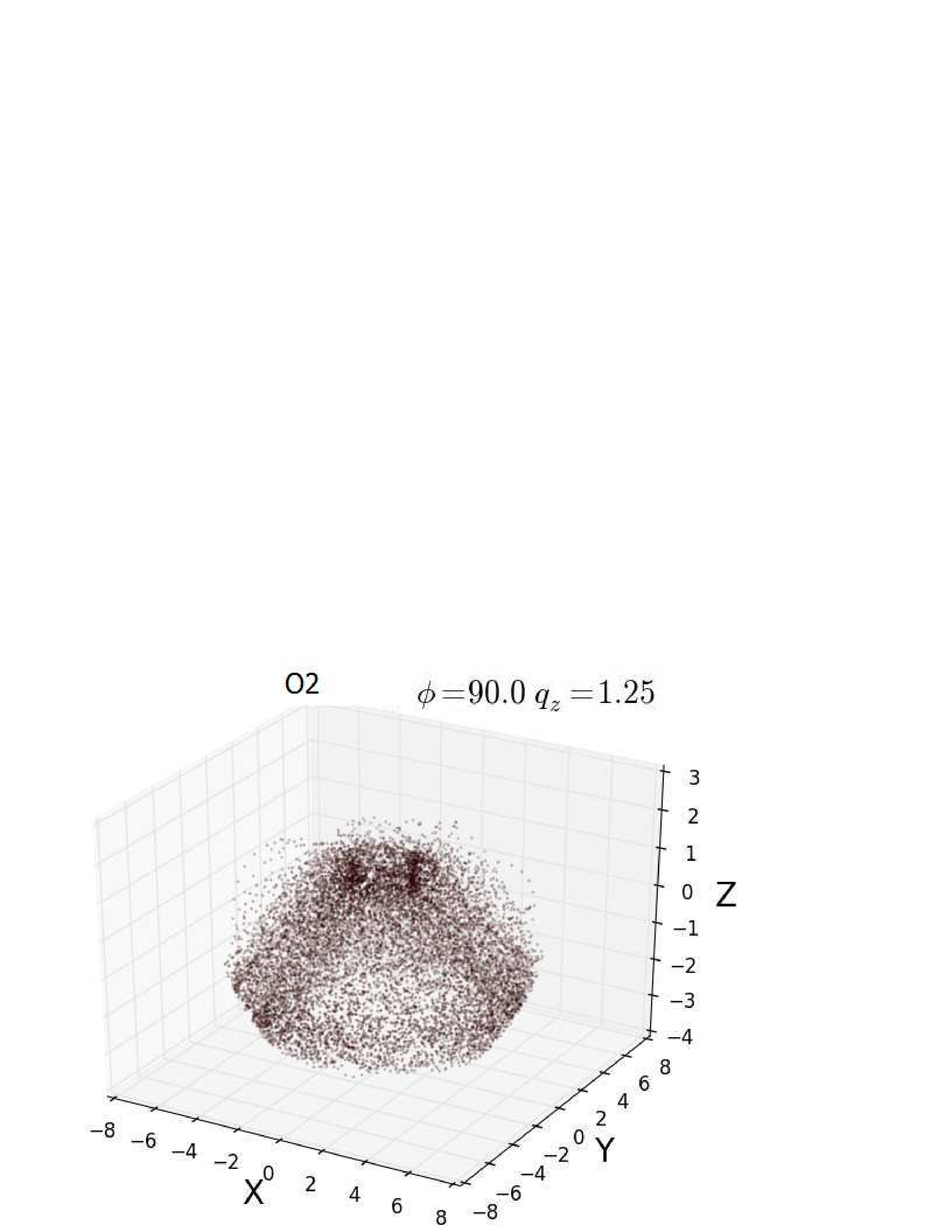} &
\includegraphics[width = 8cm]{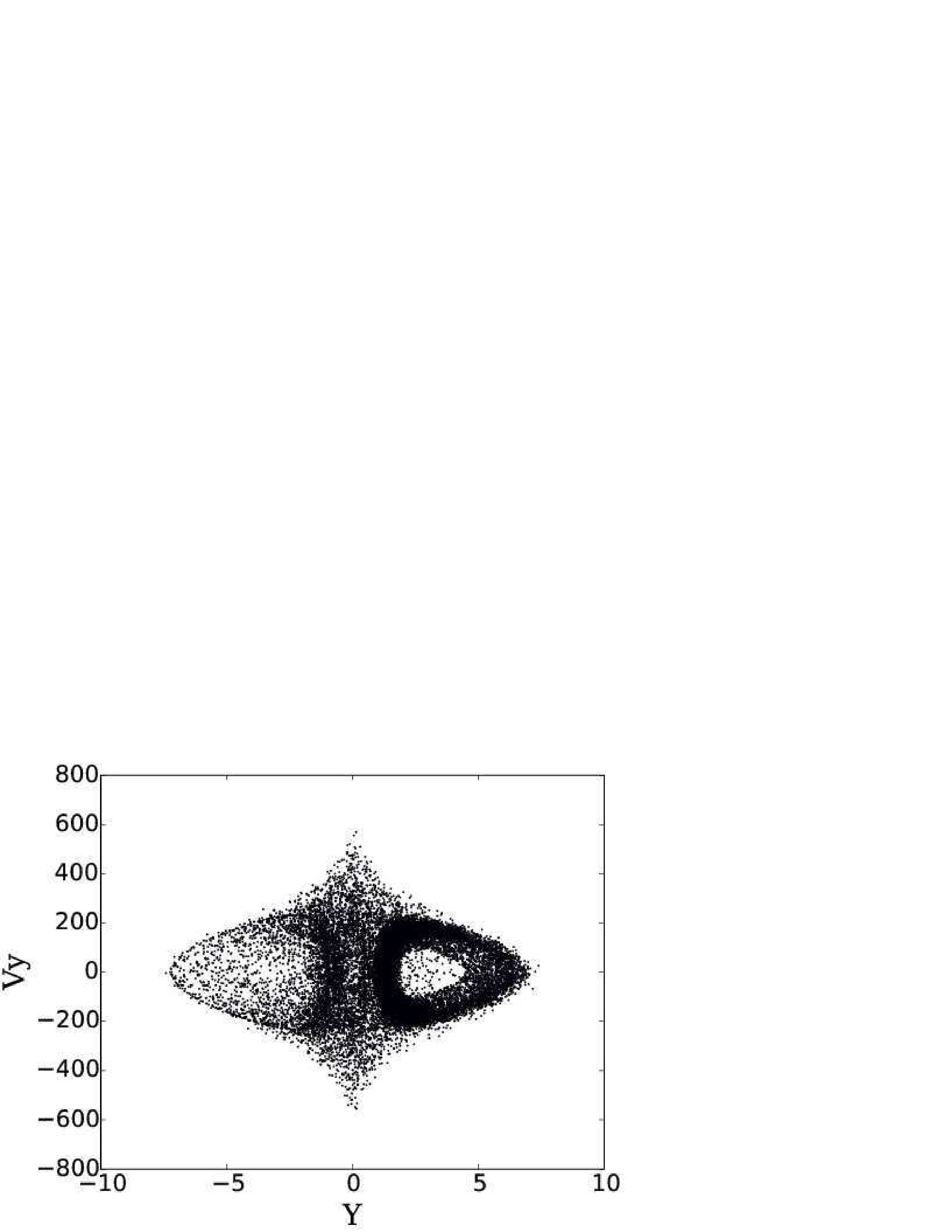}
\\
\includegraphics[width = 8cm]{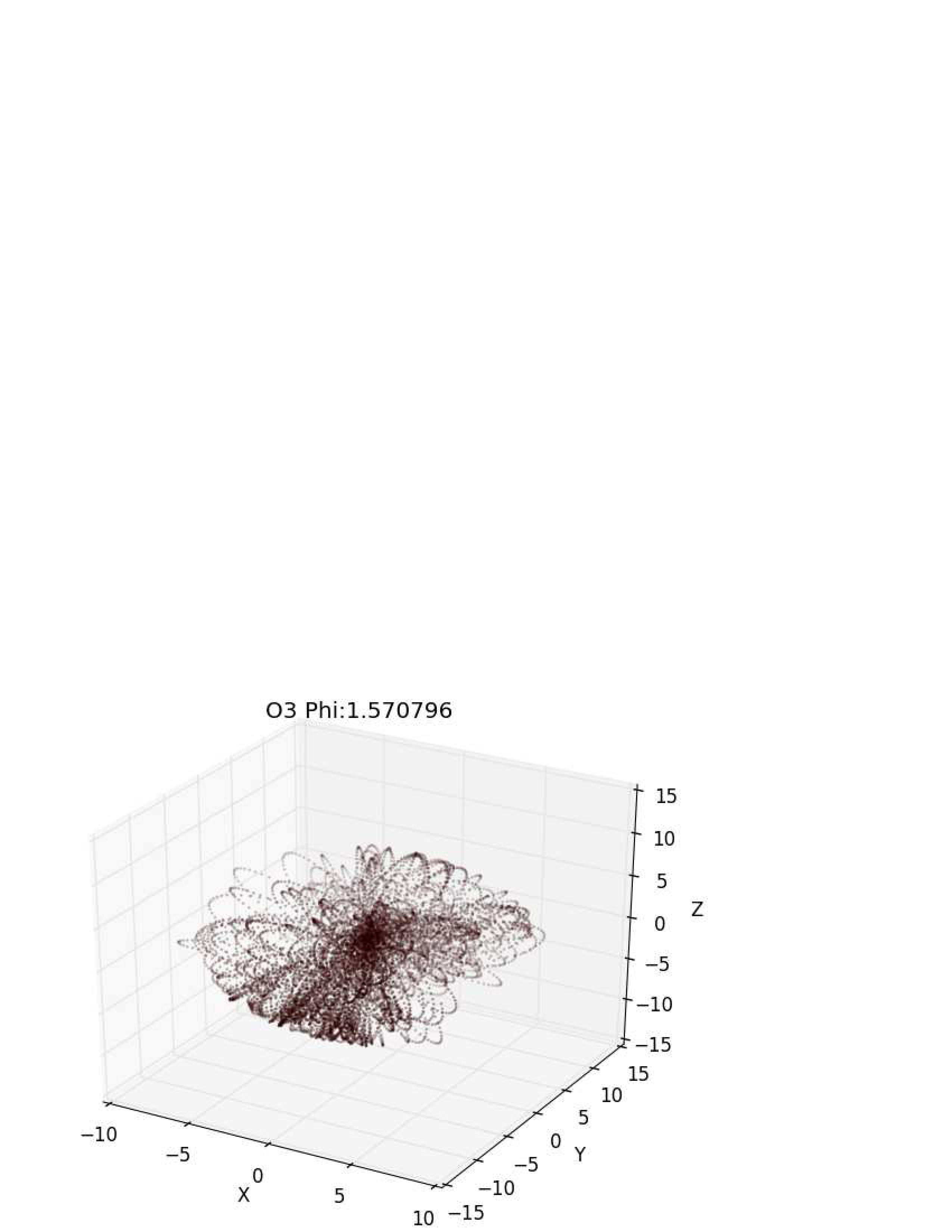} &
\includegraphics[width = 8cm]{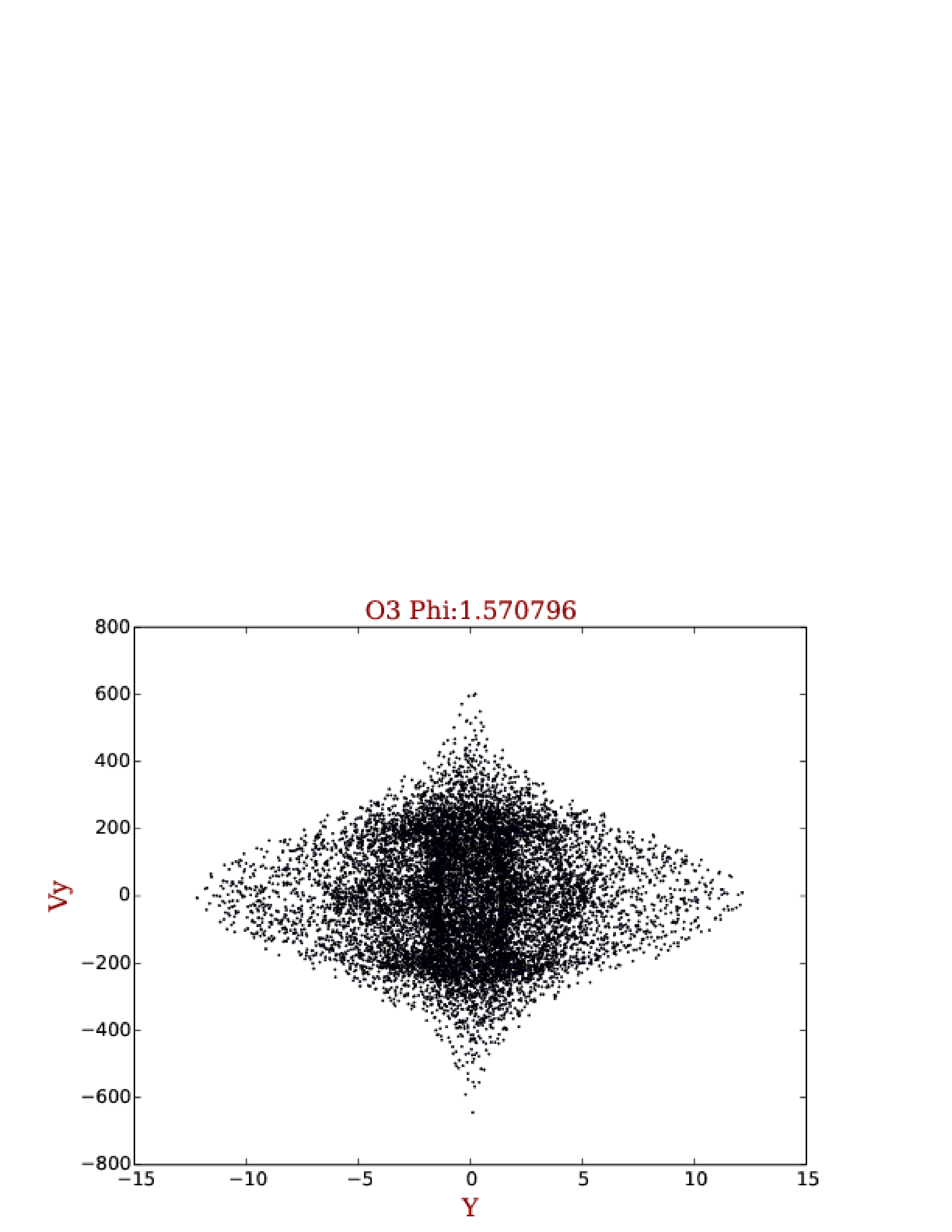}
\\
\end{tabular}
\end{center}

\caption{
Physical trajectories, and the corresponding
Poincar\'{e} sections $y - v_y$, with plane $x=0$, and $v_x > 0$,
for the selected Representative orbits, for a dark halo orientation
of $\phi = 90$.
}
\label{fig2}
\end{figure*}

\subsubsection{Lyapunov Asymptotic Exponents}

We can conclude from the inspection of these figures all orbits seem to be regular orbits,
 because they are represented
by an tori in the surface of  sections (Katsanikas \& Patsis 2011).
In order to crosscheck the results, we have computed the Lyapunov Asymptotic Exponent
for the selected orbits. The ordinary, or \emph{asymptotic} Lyapunov exponent 
can be defined as:

\begin{equation}
\label{global}
\lambda (\mathbf{x},\mathbf{v} ) =
     \lim _{t\rightarrow \infty }\: \frac{1}{t}\ln \| D \mathbf{\phi}(\mathbf{x},t) \mathbf{v} \|,
\end{equation}
provided this limit exists (Ott \& Yorke 2008). Here, $\mathbf{\phi} (\mathbf{x},t) $ denotes
the solution of the flow equation,
such that $ \mathbf{\phi}(\mathbf{x}_0,0) = \mathbf{x}_0 $,
and $D$ is the spatial derivative in the direction of an infinitesimal displacement
$\mathbf{v}$.

A system trajectory is chaotic if it shows at least one positive Lyapunov exponent,
the movement is confined within certain limited region, and the $\omega$-limit set is
not periodic neither composed of equilibrium points (Alligood et al. 1996). Conversely,
if the maximum asymptotic Lyapunov exponent is zero, this reflects the existence
of a regular motion (that is, a quasi-periodic orbit). Finally, a negative value will
reflect the existence of one attractor,
but this is not possible in a conservative system like the hamiltonian we are analysing.

The Lyapunov exponents are computed by calculating the growth rate of the
orthogonal semi-axes (equivalent to the initial deviation vectors) of one ellipse
centered at the initial position as the system evolves. By solving at the same time
the flow equation and the fundamental equation of the flow
(that is, the distortion tensor evolution), we can follow the evolution of the
vectors, or axes, along the trajectory, and in turn, their growth rate.
This method is described in (Benettin et al. 1980).

The selection of the initial deviation vectors is of importance when computing
the Lyapunov exponents during finite integrations, leading to the so-called
finite-time Lyapunov exponents (Vallejo 2003). But when one uses long enough integration
times, the axes evolve towards the fastest growing direction and the
computation of the growth rates return the asymptotic Lyapunov values (Vallejo 2008)

We have used integration of up $T = 10^5$ time-units and the resulting
maximal Lyapunov values have been nearly zero in all cases, confirming the analysis
done using the Poincare section methods, 
it is seen that all
orbits retain their regular characteristics along the stellar evolution, the regular
orbit bound to the surface of 3-dimensional torus (Kov\'{a}r e al. 2013) in the phase
space forms a narrow curve with zero width. This may be due to a different kick velocities due to SNe mass-loss and natal
kicks to the newly-formed NS (Podsiadlowski et al. 2004).
 By examining the Poincar\'{e} section in each case, we found that the system looks
 integrable
and its trajectories lie mostly on tori. These tori are represented by 2-D tori in the
surface of section. This is known the KAM
(Kolmogorov-Arnold-Moser) theorem (Kolmogorov 1954; Moser 1962; Arnold 1963).

\subsection{Orbits in the dark triaxial halo case}
To investigate the possible effects of spiral structure on the orbital characteristics of trajectory and projections, we adopt a triaxial model, the same expression as Law et al. (2009)

\begin{equation}
 \label{eq:(5)}
 \Phi_{halo}=v^{2}_{halo} ln (C_{1}x^{2}+C_{2}y^{2}+C_{3}xy+(\frac{z}{q_{z}})^{2}+r^{2}_{halo})
 \end{equation}

where \emph{v$_{halo}$} = 128~ km s$^{-1}$, \emph{q$_{z}$} represents the flattening perpendicular to the Galactic plane. The various constants $C_{1}$, $C_{2}$ and $C_{3}$ are given by

\begin{equation}
\label{eq:(6)}
C_{1}=(\frac{cos^{2}\phi}{q^{2}_{1}} + \frac{sin^{2}\phi}{q^{2}_{2}})
\end{equation}

\begin{equation}
\label{eq:(7)}
C_{2}=(\frac{cos^{2}\phi}{q^{2}_{2}} + \frac{sin^{2}\phi}{q^{2}_{1}})
\end{equation}

\begin{equation}
\label{eq:(8)}
C_{1}= 2 sin\phi~ cos\phi ~(\frac{1}{q^{2}_{1}} - \frac{1}{q^{2}_{2}})
\end{equation}
The control parameters of this model are the orientation of the
major axis of the triaxial halo $\phi$ and its flattening $q_{z}$.
(A detailed discussion on the effects of dark haloes and their role in the dynamics
of the galaxies is presented in Vallejo \& Sanjaun 2015  and references there.)

This potential is triaxial, rotated and more realistic. As consequent, it reproduces the flat rotation curve for a Milky-Way-type galaxy and it can be easily shaped to the axial ratios of the ellipsoidal isopotential surfaces (see Vallejo \& Sanjaun 2015 for details).
We include computations of the three orbits correspondence to O1, O2, and O3 which can be found in realistic galactic-type potentials that
incorporates spiral arms (see Figs. 6  \&  7).
The main parameters to play with are the flattening ($q_{z}$) and the orientation ($\phi$) of the dark halo, since the efficiency of bar formation depends very strongly on the initial orientation of the galaxy disk (Lokas et al. 2015). A lot of work on periodic orbits in the plane of a rotating barred galaxy was carried out by Contopoulos et al. 1980; Papayannopoulos \& Petrou 1983; Mulder \& Hooimeyer 1984; Zotos 2014. All integrable triaxial potentials
have a similar orbital structure (e.g. de Zeeuw 1985; Valluri \& Merritt 1998; Skokos 2001; Contopoulos 2004;
Patsis et al. 2009; Zotos \& Carpintero 2013; Patsis et al. 2014). The plots we delivered were done in spherical coordinates by two different values of the dark halo orientation ($\phi=0$ and $\phi=90$). Figs. 6\&7 ($\phi=0$, q$_{z}=0$ and $\lambda$ relatively small value and close to zero for O1) show regular orbits (tube orbits with short and long axes) for oblate and prolate cases, respectively. The effect of spiral structure is presented for all orbits and seems to be chaotic regions on the Poincar\'{e} sections (($\phi=90$, q$_{z}=1.5$ and $\lambda \geq1$ for O2 \& O3 in Figs. 6\&7).  Thus NSs in this potential follow harmonic and rotating motion in each of the $x$,$y$,$z$ directions independently. These control parameter values are given in Table 2 for all orbits. In Figs. 6 \& 7 (O1), when $\phi=0$ (stationary) $q_{1}$ is then aligned at stable equilibrium point with the Galactic $x$-axis, and the motion is stable along the it's axis. The results in this case are comparable with non-triaxial, purely logarithmic potentials. When $\phi=90$, $q_{1}$ (see Figs. 6 \& 7 for O2\&O3) is then aligned with the Galactic y-axis and it takes the role of $q_{2}$.
We note that tube tori appear in the 3D projections of the spaces of section as soon as a perturbation is introduced, even if it is a small one (Katsanikas \& Patsis 2011).
At ($\phi$, q$_{z}$) = (0,0) the outer long-axis tube in Fig. 6 is wider in the x direction than Fig. 7 when seen in
projection.
It is noteworthy to mention that Fig. 6 is symmetric in the  v$_{y}$ - $y$ plane, while Fig. 7 is not due to the Coriolis force (Gajda et al. 2016).




\section{Conclusions} \label{sec:rest}
We have found many interesting aspects for simulation of several prototypical orbits of isolated
old NSs and their connections with the spatial distribution
phenomena, through the long-term stellar dynamics. We employed 3-D Galactic
gravitational potential models of  normal non-barred galaxies.
The models consist of axisymmetric and a triaxial dark halo potentials with three components:
bulge, disc, and dark halo. These are relatively
simple potentials that can show complex behaviors, which
are found in more realistic galactic-type potentials.
Then we found that orbits can rotate around the axis
 of symmetry of the phase space on a surface of section, in the same direction in 3-D and 2-D loops for various values of the initial conditions for the family of axisymmetric and triaxial models. We have used the Poincar\'{e} technique to study the 3-D NS trajectories and their 2-D projections. It is shown that both loop and family of orbits arise as a natural consequence of the dynamics of the stable periodic orbits (with regular motion). In addition, the morphology of orbits in triaxial potentials can determine the structure of triaxial galaxy itself.
There are 3-D invariant tori containing quasi-periodic motions,
these tori are represented by 2-D tori on the surface of the section associated with phase space.
In addition, the Lyapunov asymptotic exponent is equal to zero in case of non-triaxial.  However, it is clear that the chaotic orbit
 have a non-zero real exponent, since the finite-time Lyapunov exponents distributions reflect the underlying dynamics
(Vallejo et al. 2003). We also show that the phase space in triaxial galaxies with a rotated halo is rich in regular and chaotic regions, which is consistent with the analysis done by the Poincar\'{e} cross sections.
The results of these motions are strongly
affected by the gravitational potential of the galactic disk
associated with the effect of kick velocities on the orbital
parameters. This can provide us a better
visualization of the old NS dynamics.

The conclusions of the present research
are considered as an initial effort and also as a promising
step in the task of understanding  the underlying dynamics
of the phase space by mapping the regular
regimes in 3-D axisymmetric potential. 
Future work will go steps further in detection
of the old NSs via accretion of the interstellar medium material
which may make them shine, and their weak luminosity could be detected as
soft X-ray sources ($0.5-2$ keV). However,  $\emph{eROSITA}$ all-sky survey have an
excellent capabilities of
available survey for this kind of studies, and it should improve our
knowledge of Galactic NSs phenomenally for the next decade. $\emph{eROSITA}$ will be about
 20
times more sensitive than the ROSAT all sky survey in the soft X-ray band (Merloni et al.
2012).

\section*{Acknowledgments}

We specifically would like to acknowledge Ying-Chun Wei for his assistance with the
simulation code.
We are very grateful to Haris Skokos, Panos Patsis, Matthaios Katsanikas, Georgios
Contopoulos, Nicola Sartore, Matthew Molloy, Diomar Cesar Lob\^{a}o, Barbara Pichardo
and Mar\'{\i}a de los Angeles Perez Villegas for fruitful discussions and useful
remarks. The authors would like to thank the anonymous referees
for the careful reading of the manuscript and for all suggestions and comments which
allowed us to improve both the quality and the clarity of the paper.


\end{document}